\documentclass[11pt,preprint]{aastex}
\usepackage{epsf,epsfig}

\newcommand{\flux}{\mbox{erg~s$^{-1}$~cm$^{-2}$}}
\newcommand{\lum}{\mbox{erg~s$^{-1}$}}
\newcommand{\Msun}{\mbox{M$_\odot$}}

\begin{document}
\shorttitle{MS0451.6-0305}
\shortauthors{Donahue et al.}

\title{The Mass, Baryonic Fraction, and X-ray Temperature of the 
Luminous, High Redshift Cluster of Galaxies,
MS0451.6-0305}
\author{Megan Donahue}
\affil{Space Telescope Science Institute, 3700 San Martin Drive,
Baltimore, MD, 21218, donahue@stsci.edu}
\author{Jessica A. Gaskin, Sandeep K. Patel\altaffilmark{1} and Marshall Joy} 
\affil{Department of Space Science, SD50 \\ 
        NSSTC/NASA Marshall Space Flight Center \\ 
        Huntsville, AL 35805}
\altaffiltext{1}{National Research Council Associate} 
\author{Doug Clowe}
\affil{Institut f\"ur Astrophysik und Extraterrestrische Forschung
der Universit\"at Bonn, Auf dem H\"ugel 71, D-53121 Bonn,
clowe@astro.uni-bonn.de}
\author{John P. Hughes}
\affil{Department of Physics and Astronomy, Rutgers, The State
University of New Jersey, jph@physics.rutgers.edu}

\begin{abstract}
We present new Chandra X-ray observations of the luminous and
cosmologically-significant X-ray
cluster of galaxies, MS0451.6-0305, at $z=0.5386$. 
Spectral imaging data for the cluster are consistent with an
isothermal cluster of $(10.0 - 10.6) \pm1.6$ keV, with an intracluster
Fe abundance of $(0.32-0.40) \pm 0.13$ solar. 
The systematic
uncertainties, arising from calibration and model
uncertainties, of the temperature determination 
are nearly the same size as the statistical 
uncertainties, since the time-dependent correction for 
absorption on the detector is uncertain for these data. We
discuss the effects of this correction on the spectral
fitting. The effects of statistics and fitting assumptions
of 2-D models for the X-ray surface brightness 
are thoroughly explored. This cluster
appears to be elongated and so we quantify  
the effects of assuming an ellipsoidal gas distribution on 
the gas mass and the total gravitating mass estimates.
These data are also jointly
fit with previous Sunyaev-Zel'dovich observations to
obtain an estimate of the cluster's  
distance ($D_A= 1219^{+340}_{-288} \pm 387$ Mpc, statistical 
followed by systematic uncertainties) assuming spherical 
symmetry. If we, instead, assume
a Hubble constant, the X-ray and S-Z data 
are used together to test the consistency of an ellipsoidal
gas distribution and to weakly constrain the intrinsic axis
ratio. The mass derived from the
X-ray data is consistent with the weak lensing mass and is only marginally less
than the mass determined from the 
optical velocities. We confirm that this cluster is very hot and
massive, further supporting the conclusion of previous analyses that
the universe has a low matter density and that cluster properties have
not evolved much since $z\sim0.5$. Furthermore the presence of
iron in this high redshift cluster at an abundance that is the same as
that of low redshift clusters implies that there has been very
little evolution of the cluster iron abundance since $z\sim0.5$.
We discuss the possible detection of a faint, soft, extended component that may be
the by-product of hierarchical structure formation.
\end{abstract}


\section{Introduction}

Clusters present a rich source of cosmological data. If all massive 
clusters of galaxies are indeed fair samples of the universe,  
observations even of individual clusters can yield estimates of the 
 baryon to dark matter ratio and mass to light ratio of the Universe
 (e.g.\ Carlberg, Yee, \& Ellingson 1997; Allen, Schmidt \& Fabian 2002).
Furthermore, 
rate of formation of the largest gravitationally-bound structures 
in the universe, which can be estimated from the evolution of 
the cluster mass function, is 
sensitive to the mean density of the universe, or $\Omega_M$
(e.g.\ Eke, Cole, \& Frenk 1996).  
Studies that include 
sufficiently distant clusters could also constrain the acceleration
parameter, or $\Omega_\Lambda$ (Haiman, Mohr, \& Holder 2001). 
A recent review of clusters and
cosmology can be found in Rosati, Borgani, \& Norman (2002), 
(See also Schueker et al (2003) and references therein for results 
from the ROSAT ESO Flux Limited (REFLEX) X-ray cluster sample.) 

Even in the era of cosmology after the spectacular success of
WMAP (Wilkinson Microwave Anisotropy Probe; Bennett et~al.\ 2003), 
other methods of estimating 
cosmological
parameters provide independent tests of microwave background 
results. Indeed, even WMAP uses external constraints as priors
to achieve its most precise results (Spergel et~al.\ 2003). 
Constraints from clusters of
galaxies are orthogonal to those from supernovae studies and from
the microwave background; the constraint from the evolution of the
mass function is independent of the Hubble constant. Furthermore,
cosmological studies of clusters test the physics that underlie the
assumptions: the physics of gravitational formation of structure and
its effects on baryonic matter. WMAP gave us an unprecedented picture
of the universe at the time of recombination; studies of clusters 
can tell us how well we understand the laws of physics that we 
assume to predict the current universe, based on what we have seen
at recombination.

X-ray cluster samples, such as that drawn from the  
Extended Medium Sensitivity Survey (EMSS, see Gioia et~al.\ 1992ab), 
have been used to place statistical constraints
on the baryon fraction of the universe (e.g.\ Lewis et~al.\ 1999) 
and on the evolution of 
the cluster temperature function (Eke et~al.\ 1998; 
Donahue \& Voit 1999; Henry 2000). 
In their normalization and behavior as a function of redshift, 
the distributions of 
cluster luminosity, X-ray temperature, and in particular, 
mass are very sensitive to 
cosmological parameters such as $\Omega_M$ and $\sigma_8$
(White, Efstathiou, \& Frenk 1993; Eke, Cole, \& Frenk 1996). Since the evolution of
the cluster temperature function is greater for the hotter and more massive clusters, 
the estimate of $\Omega_M$ is most sensitive to the evolution of the rarest and
most massive systems. However, the
power of all such observations depends on how robustly one can infer the 
cluster mass based on observations of cluster X-ray temperature and
temperature gradients. The intracluster medium of a 
cluster is usually assumed to be 
 in hydrostatic equilibrium, nearly spherical, and quite smooth. 
Until the advent of the Chandra X-ray Observatory mission, X-ray observations
did not have the combined spatial and spectral resolution required to prove
otherwise. Phenomena such as major shock fronts would be smeared out by the
poor resolution of the X-ray telescope or the detectors. 

In order to measure the mass and hot baryonic component of a 
distant, hot, and therefore cosmologically important 
cluster, we observed one of the most
luminous, hot, and distant  
clusters in the EMSS, MS0451.6-0305.  
Our goals were to revisit previous X-ray spectral estimates of
the temperature, gas mass, metallicity,  
and, if possible, to detect the presence of temperature and 
metallicity gradients and to 
infer the intrinsic geometry of the cluster.  In combination with
$H_0$ constraints from other experiments, we also  
jointly fit the Chandra observations with previous Sunyaev-Zel'dovich 
observations to estimate the intrinsic geometry and obtain consistent estimates
of the gas mass and total mass. 

MS0451.6-0305 is an extremely luminous X-ray cluster at $z=0.5386$. It 
was serendipitously detected as an X-ray source with the EINSTEIN Imaging
Proportional Counter, and it  was
identified as a cluster in the Extended Medium Sensitivity Survey 
(EMSS) by Stocke et~al.\ (1991).
Its high redshift was confirmed by Gioia \& Luppino (1994); a more refined
central redshift was obtained by the Canadian Network for Observational
Cosmology (CNOC) (Yee, Ellingson, Carlberg 1996). X-ray follow-up was 
obtained  
 with ROSAT (Donahue, Stocke \& Gioia 1994) and ASCA (Donahue 1996;
Donahue et~al.\ 1999). MS0451.6-0305 is one of the most luminous clusters
in the EMSS ($14.6 \times 10^{44} h^{-2}$ erg s$^{-1}$ (0.3-3.5 keV)), 
with an X-ray temperature ($\sim10$ keV) and velocity
dispersion (1350 km s$^{-1}$ (Carlberg, Yee, \& Ellingson 1997)) to match.
Based on these observables, along with weak lensing constraints
on the mass by Clowe et~al.\ (2000), 
this cluster is thought to be a very massive cluster, and therefore very 
important for cosmological cluster studies. The observations  
described in this paper for this
cluster are listed in Table~\ref{tab:obs}. We also review optical results from
Carlberg et~al.\ (1996) and weak lensing results from Clowe et~al.\ (2000).

In Section 2, we present the Chandra observations and data reduction. In Section
3, we discuss our X-ray data analyses of the spectra and the X-ray image. In
Section 4, we present the pedagogical basis for inferring gas densities, gas 
masses, and gravitational masses from X-ray data for both spherical and 
ellipsoidal shape assumptions. In Section 5 we apply those relations to our 
data and compare our results to mass measurements from the literature. 
In Section 6 we report the results of jointly fitting the X-ray data with
Sunyaev-Zel'dovich observations, and in Section 7 we discuss our conclusions.
For this paper, we assume a flat cosmology with $\Omega_M=0.3$ and $H_0=
100 h$ km s$^{-1}$ Mpc$^{-1}$ unless otherwise stated. This cosmology is
compatible with WMAP results if $h=0.7$. The angular scale
at $z=0.5386$ is $4.44 h^{-1}$ kpc$/\arcsec$.

\section{Chandra Observations and Data Reductions}

MS0451.6-0305 was observed with
the Chandra Advanced CCD Imaging Spectrometer (ACIS) detector on 8-9 Oct 2000 
for a total of 44,192 seconds (OBSID 902). The ACIS-S3 light curve was inspected to identify the time period 
of high background rate for a single flare which exceeded
a 10\% threshold above the mean. This time period was identified and
removed from the data to result in a net 41,248 seconds of exposure
time.  The observed,
background-subtracted 
count rate for 0.7-7.0 keV was 0.2974 counts sec$^{-1}$ in an
aperture of 163 pixels ($80.2\arcsec$) 
in radius and a background rate of 0.027 counts sec$^{-1}$.
The cluster was centered on ACIS chip S3, a backside-illuminated
device with good charge transfer efficiency (Townsley et~al.\ 2000) 
and spectral resolution; data were acquired with the timed 
exposure mode and FAINT mode options. 

All of the processing, calibration, 
and much of the analysis and extraction of the X-ray
data were done with {\em Chandra Interactive Analysis of 
Observations} (CIAO) v2.2.1 and v2.2.3 
packages\footnote{http://cxc.harvard.edu/ciao}
along with the calibration database (CALDB) 
available from the Chandra X-ray Center (CXC); 
spectra were analyzed with the 
XSPEC\footnote{http://heasarc.gsfc.nasa.gov/docs/xanadu/xspec/index.html} v11.1 (Arnaud 1996). For this paper, the names of the CIAO packages
will be italicized. Since the interpretation of X-ray data  
depends on the maturity of the calibration, 
we report version numbers when available and release dates
otherwise. 

The name of the 
original processing implemented by the CXC was R4CU5UPD11, so 
we reprocessed the Level 1 event files with the
gain files from CALDB v2.10 using {\em acis\_process\_events} with the 
appropriate bad pixel files. The level 1 events
were filtered with the standard good time intervals supplied by 
the pipeline, and then filtered to admit only ASCA grades 0, 2, 3, 4, and
6, and ``clean'' status ($=0$) events. The plate scale for the
unbinned data is $0\farcs4920$ per detector  
pixel. The aspect for this observation
required a very small correction of $\Delta RA = -0.63\arcsec$ 
and $\Delta Dec = -0.58\arcsec$. A typical absolute astrometric uncertainty
in a Chandra ACIS-S observation is about $1.0\arcsec$ but it can be as large as
$3.0\arcsec$ relative to astrometric standards in the International Celestial
Reference Frame (ICRS) and Hipparcos (the Tycho2 catalog).\footnote
{Chandra Positional Accuracy Monitor,
http://cxc.harvard.edu/cal/ASPECT/celmon/index.html.}

Spectral analysis was performed in Pulse Invariant (PI) space
(i.e., after the instrument gains were applied) using the gain map
appropriate for the focal
plane operating temperature on the dates of these observations, $T_{fp}=-120$C.
The deep background file acisD2000-08-12gainN0003.fits (available in the
public Chandra calibration database CALDB) was reprocessed
to use the gain file consistent with the data. The background 
data, as supplied by the CXC in Feb, 2002, 
were reprojected
using the aspect solution for our observation. The 
count rates from the deep background files were similar to those in our 
observation to better than 1\% over the energy range of interest to us
$0.7-7.0$ keV. We used these files to model the background 
for both the spectral and the spatial analyses.

We discuss the application of a soft-energy, time-dependent correction
to the throughput of the ACIS-S. We report results with and without
this time-dependent correction, which affects the calibration
mainly at energies less than one keV. The correction is applied
to the ``arf'' (area response file), which is used in conjunction with the ``rmf" (redistribution matrix file) to model telescope and instrument response.\footnote{
An IDL program, ${\rm acisabs.pro \, v}1.1$, was
provided by George Chartas of Penn State University to modify the arf
files. The URL for this package is 
http://www.astro.psu.edu/users/chartas/xcontdir/acisabsv1.1\_idl.tar.gz.} 

In preparation for inspection and analysis of the spatial data, 
we generated two exposure maps to correct approximately 
for the variation in sensitivity and in net observing time across the image.
Since most of the source photons are relatively soft, we assumed 
a monochromatic X-ray spectrum of 1 keV photons. 
For the 2-D spatial fitting, the map was matched to the fit  
data by binning pixels $8 \times 8$.  To generate an adaptively
smoothed, exposure-corrected 
image at the highest resolution allowed by the data,
we produced an unbinned exposure map of the central  
$512 \times 512$ pixel region of the S3 chip. Division of the image
data  by the exposure map converts the raw 
counts data (counts pixel$^{-1}$) to units of photons cm$^{-2}$ s$^{-1}$ pixel$^{-1}$.

\section{X-ray Data Analysis}

This section presents the X-ray spectra and the surface brightness maps extracted from the
Chandra data. We show that the X-ray spectra between 0.7-7.0 keV are adequately 
represented by an isothermal
cluster, and we have no evidence 
for significant temperature fluctuations in the region studied. 
However, depending on the application of 
the time-dependent soft-energy correction, we also see
evidence for extended, faint, soft  X-ray 
emission ($\lesssim 1$ keV). 
The spectral shape of the soft component is equally well fit
by cool thermal X-ray plasma or a steep power law.
This emission could be associated with groups
falling into the cluster. Some evidence for such gas may be visible in
the softest X-ray map for the cluster, which has a different centroid
and peak location 
than the hard X-ray maps. We will also show 
the cluster X-ray surface brightness profile is best represented 
by an elliptical ``beta'' model. 

\subsection{Cluster Temperature and Metallicity}

\subsubsection{Global Temperature}

Previous Chandra analyses of distant clusters (e.g.\ Jeltema et~al.\ 2001) 
illustrated the need to mask point source emission contaminating cluster
spectra. Chandra's spatial resolution makes the identification of point
sources comparatively straightforward. 
The raw Chandra data for MS0451.6-0305 show only two faint point sources within
$1\arcmin$ of the cluster center. 
Molnar et~al.\ (2002), in a study of the statistics of X-ray point sources
around this cluster, identify these sources at
04h 54m 12.81s, $-03\degr \, 00\arcmin \, 47.7\arcsec$
and
04h 54m 10.88s, $-03\degr \, 01\arcmin \, 25.2\arcsec$. They are both 
very
soft sources with no detected counts $>2$ keV, and only $45.2 \pm 8.1$
counts and $21.4 \pm 5.7$ counts 
between 0.5--2.0 keV, respectively. Neither of these sources has
sufficient counts to affect the spectral or imaging analyses in this
paper.

We extract the X-ray events from a circle centered on
RA 04h 54m 10.80s and Dec $-03\degr \,  00\arcmin \, 51\farcs8 $ 
(J2000), 168 detector pixels in radius ($r=83\arcsec$).
We masked 3 possible point sources, the two mentioned above and a third
even fainter source, but their contamination 
represented a very small contribution of counts to a spectrum of $\sim13,000$
counts. 
We binned the spectrum to a minimum of 20 counts per energy bin in the combined
source and background spectrum. We extracted the events from an identically
filtered version of the deep background field provided by the
Chandra X-ray Center.\footnote{http://cxc.harvard.edu/ciao/threads/acisbackground.} 
Within 0.7--7.0 keV, 91.7\% of the 13,383 X-ray 
events were source counts. A weighted photon redistribution matrix
file (rmf) and weighted area response file (arf) were created using
the prescription of the Chandra Science Center for CIAO2.2.1.\footnote
{http://cxc.harvard.edu/ciao2.2.1/threads/wresp\_multiple\_sources/}  

\begin{figure}
\includegraphics[angle=270,width=\textwidth]{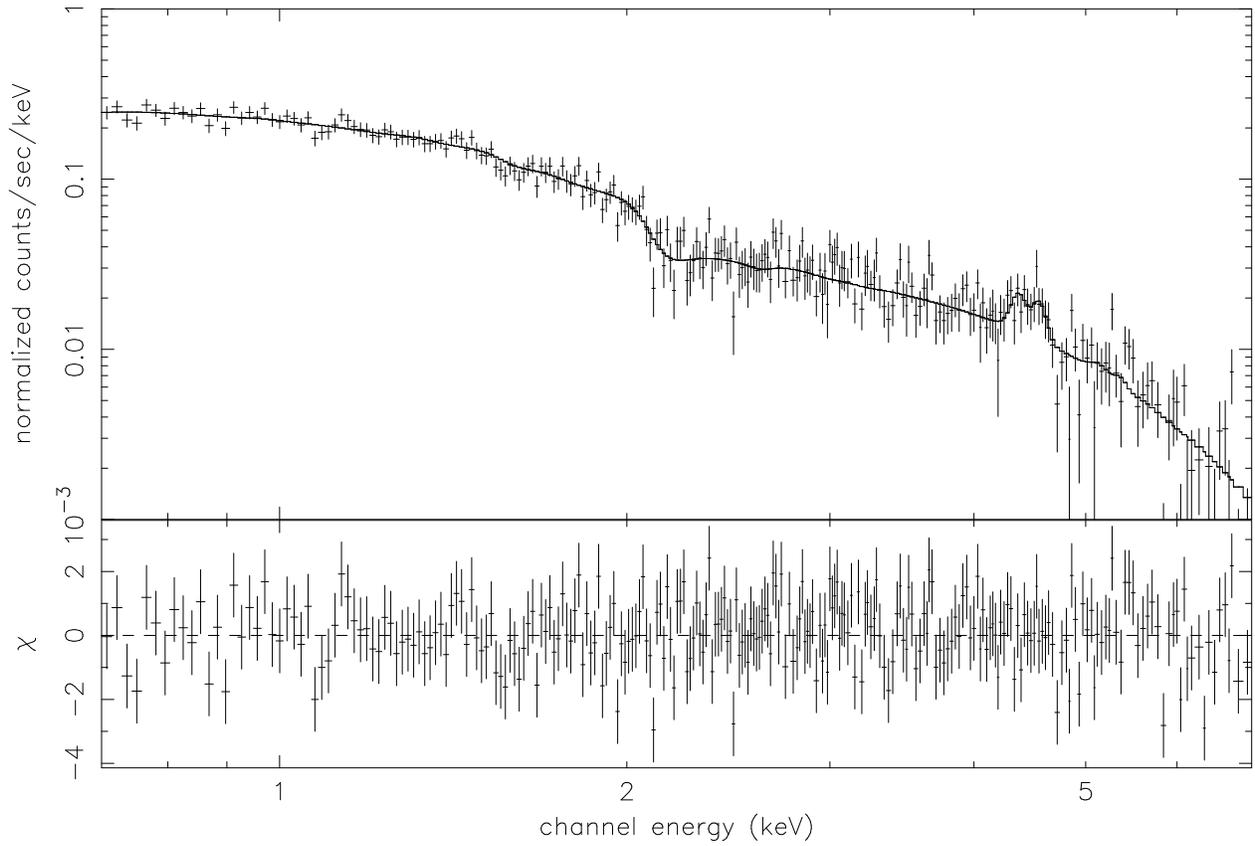}
\caption{The full Chandra spectrum for MS0451.6-0321, with the response function
left in, as a function of energy (keV). The solid line is the best fit thermal spectrum to the data. 
The second window shows the $\Delta \chi^2$ for the fit, as a function of energy. 
\label{figure:spec}}
\end{figure}

We first discuss the spectral analysis of the data uncorrected 
for a time-dependent absorption feature on the ACIS-S detector.
The extracted spectrum (Figure~\ref{figure:spec}) was fit using 
XSPEC (v11.2). We
used  MekaL and Raymond-Smith models with Galactic absorption.
Relative metallicities were assumed to be the meteoritic 
abundances from Anders \& Grevesse (1989).
In particular, the solar iron abundance was assumed to be
meteoritic, or $4.68 \times 10^{-5}$
solar (Anders \& Grevesse 1989). 
We found an excellent fit to either model, with very little difference
in the best fit or uncertainty ranges for each (Table~\ref{global}). 
The best fit temperature 
changes somewhat if we fix the Galactic $N_H$ value at 
$5.0 \times 10^{20}$ cm$^{-2}$ (Table~\ref{global}.) 
Galactic $N_{HI}$ is $4.7-5.2 \times 10^{20}$ cm$^{-2}$ in the
direction of the cluster (Dickey \& Lockman
1990; Stark et al 1992.) Table~\ref{global} lists the reduced
$\chi^2$ and the probability (``Prob'') of finding a larger 
$\chi^2$ value given the data uncertainties if the model is correct.

The best-fit results and goodness-of-fit assessments 
are sensitive to the chosen bandpass (Table~\ref{bin}). 
Fits including the energy range $<0.6$ keV have significantly higher $\chi^2_{red}$
than fits restricted to higher energy ranges. No improvements to the fit
at this lower energy range are made by including temperature or
absorption components in the model. To avoid  sensitivity
to the soft energy calibration and possible background problems,
we restricted our fits to $E>0.7$ keV for all of our final 
spectral analyses. The errors we quote for each measurement 
are the formal, statistical 90\% confidence
range for a single interesting parameter ($\Delta \chi^2 = 2.7$).
The systematic uncertainties based on these 
fitting choices are arguably about the same
size as the statistical uncertainties for such a hot cluster if
no restriction on the fitted bandpass is included.
The Chandra global temperature estimates for MS0451.6-0305 are  
consistent with the ASCA temperature of $10.9\pm1.2$ keV 
(Donahue, 1996; Donahue et~al.\ 1999) even though 
the Chandra extraction aperture size
of radius $r=163$ pixels ($\sim80\arcsec$) 
is significantly smaller than that used 
for the ASCA observations ($6.0\arcmin$). 

The cluster iron abundance estimate is relatively independent of
the technique used to fit the data, $\sim35-40\%$ solar. 
As we will discuss in \S3.1.2, the statistics are not sufficient to determine 
whether a metallicity gradient exists for this cluster.

We investigated the impact of the time-dependent 
soft energy correction 
released by the CXC on July 29, 2002, which is applied to 
the ``arf'' file in advance of fitting the data. At the
epoch of our observation, the correction is
mainly limited to the soft bandpass ($E<1.0$ keV). The 
amplitude of the recommended correction is about twice the dispersion of
the measurement of the on-board calibration source $^{55}$Fe
L-complex/Mn-K ratio\footnote{http://cxc.harvard.edu/cal/Acis/Cal\_prods/qeDeg/}, so the magnitude of the correction is very nearly 
the same magnitude of the uncertainty of the correction. 

Including this correction 
in our analysis gave interesting results. The fit to the global spectrum, with the correction, required a second component 
which could be either a second, cooler, thermal component or a steep, power-law
component (Table~\ref{tab:soft}). The second component is a minority 
of the 0.7-7.0 keV
spectrum (contributing between $1-5\%$ of the flux in the 0.7-7.0 observed bandpass). 

The main effects of the soft energy correction are the following (see 
also Table~\ref{tab:soft}):
\begin{enumerate}
\item Models for a single temperature plasma, with an absorbing column that was allowed
to vary, settled on a best-fit temperature of $10.2\pm^{0.9}_{1.0}$ keV depending on bandpass,
as before. This temperature is 
consistent with our estimates from uncorrected spectra. However, the best-fit absorption column for
this model is consistent with zero, and inconsistent with $N_H$ expected
from 21-cm data (Dickey \& Lockman 1990; Stark et~al.\ 1992). 
Since we know we are looking through
our own Galaxy, this model is probably unphysical.

\item Fixing the Galactic column density towards MS0451.6-0305 to $5 \times 10^{20}$ cm$^{-2}$ (Dickey \& Lockman 1990; Stark et~al.\ 1992) resulted
in unacceptable single temperature 
fits, ruled out at the $98\% - 99.6\%$ confidence level,
depending on the bandpass fit. The 
best-fit temperature was $\sim7.9-7.5$ keV in these models.

\item Allowing a two-temperature plasma resulted in an acceptable fit with or 
without fixing the Galactic column. 
The hot plasma dominates the emission, with a best-fit temperature of
$10.6 \pm ^{1.5}_{1.2}$. The best-fit cool temperature is $\sim0.8$
keV when the energy range of 0.7-7.0 keV is fit. That best-fit temperature
of the cooler component is sensitive to the fitted 
energy range, increasing to $\sim2$ keV when 0.5-7.0
keV is fit. The hotter component is consistently fit
to a temperature 
between $10.2-10.8$ keV.

\item Instead of a cool component, allowing a soft component with a steep power-law spectrum with a photon index of $\sim 2.5$ 
plus hot component with a $\sim10$ keV thermal spectrum fit the data 
equally well. The data do not constrain the spectral shape of the soft
component well enough to distinguish between a power law and a thermal
spectrum.

\end{enumerate}

This analysis may also explain the discrepancy between our best-fit temperature
for this cluster and that obtained by Vikhlinin et~al.\ (2002) of $8.1\pm0.8$ keV
for the same Chandra data. We obtain best-fit values of the global temperature for the 
absorption corrected spectrum near $8-9$ keV if we fix Galactic $N_H$
and restrict the fit to $0.8-7.0$ keV. But as
we discussed above, if we allow Galactic $N_H$ to be free, we are required
to include a second, soft component, otherwise 
the implied Galactic $N_H$ is unphysically tiny. 

Most interestingly, this result implies that if the time-dependent, soft-energy correction is
correct, there is a hint of a soft excess in our data that could be 
coming from a ``cool'' component of $0.7-2.0$ keV or a 
power-law component. We could not place 
any useful temperature constraints on this cool component except that it is hot enough to be detected in the X-rays. If it is real,
it is contributing very little ($\lesssim 1-5\%)$ of the emission at $E>0.7$ keV, and 
therefore does not affect the analysis elsewhere in this paper, which
is limited to $E>0.7$ keV. But any soft excess 
is intriguing, since it could indicate the presence of infalling
material or non-thermal physical processes (Table~\ref{tab:soft}).

The energy dependence of the global temperature fit was presaged in a 
theoretical paper by Mathiesen \& Evrard (2001). They find that the spectral
fit temperature is usually lower than the mass-weighted average temperature
in their models, because of the influence of cooler gas accreted in the
expected hierarchical cluster process. We unintentionally followed
their prescription for improving the spectral estimate of the mass-weighted
temperature by restricting the bandpass to higher energies. Some other 
theoretical work has been done to predict the contribution of a warm
component of substructure to the hotter cluster halo (e.g.\ Cheng 2002); it
would be interesting to confirm and characterize this component with higher quality soft X-ray
data.

\subsubsection{Temperature and Metallicity Gradients}

In order to quantify the presence of any temperature gradients, 
we divided the photons into an inner circular aperture of $r<31\arcsec$ 
and an outer annulus of $31\arcsec - 84\arcsec$. The inner
aperture had 6,055 net counts; the outer 6,268 net counts. The extracted
spectra were binned to a minimum count per energy bin of 20 counts,
before background subtraction. The inner bin contained 97.6\% source counts;
the outer, 86.3\% source counts. Weighted rmf and arf files were
constructed as before.

When the same models and fitting constraints were placed on both fits, the 
results were very similar statistically, because the 1-$\sigma$ 
boundaries for one interesting parameter ($\Delta \chi^2 = 2.7$) were
large (see Table~\ref{inner_outer}.) For all these fits, the
column density was allowed to vary and usually found a best fit
value of $4-5 \pm 2-3 \times 10^{20}$ cm$^{-2}$. This value is 
consistent with the  
nominal Galactic column density towards MS0451.6-0305 
of about $5.0 \times 10^{20}$ cm$^{-2}$ (Stark et~al.\ 1992; Dickey \& Lockman 1990). 

The spectral 
residuals are not concentrated at low or high energies. Adding an
intrinsic $N_H$ component does not improve the fit; a best-fit intrinsic 
hydrogen absorption column density at the redshift of the cluster 
is statistically consistent with zero.

We also fit the annuli's spectra with 
the recently computed correction at the soft-X-ray
band included in the arf file. 
The poorer statistics of the divided data made constraining 
the nature of the soft component even more difficult, so we do not 
report the results in detail, since the bottom line is the same:
the model which fits the spectrum of the inner annulus is statistically
consistent, aside from overall normalization, 
with that which fits the spectrum of the outer annulus. 
A corollary of that result is that the
need for some soft component persisted in both the inner and outer
spectrum. The persistence in both spectra suggests that if the soft
component is real and not an artifact of calibration or background
 subtraction, it is extended, and it is not confined to the core
of the cluster.

There were no statistical differences in the temperature and the 
metallicity between the inner and outer apertures. However, the strength of
this conclusion is limited by the counting statistics of the data. The
difference would have had to be $\sim2$ keV to be discernable in these data for 
a cluster this hot. The data are consistent
with the cluster being nearly isothermal (at the 20\% level) out to
$r=84\arcsec$. 

We extracted a spectrum from an ellipse centered on the medium-energy
surface brightness peak that may be associated with the cluster's brightest
galaxy. (See next section.) However, the temperature is not well constrained 
by the 1867 counts extracted from the region. The best fit temperature obtained 
from the spectrum  was $10.7^{+5.8}_{-2.7}$ keV ($1\sigma$ statistical
uncertainties), and therefore was not statistically different from the rest of 
the gas.

\subsection{Cluster Morphology}

\begin{figure}[ht]
\includegraphics[angle=0,width=\textwidth]{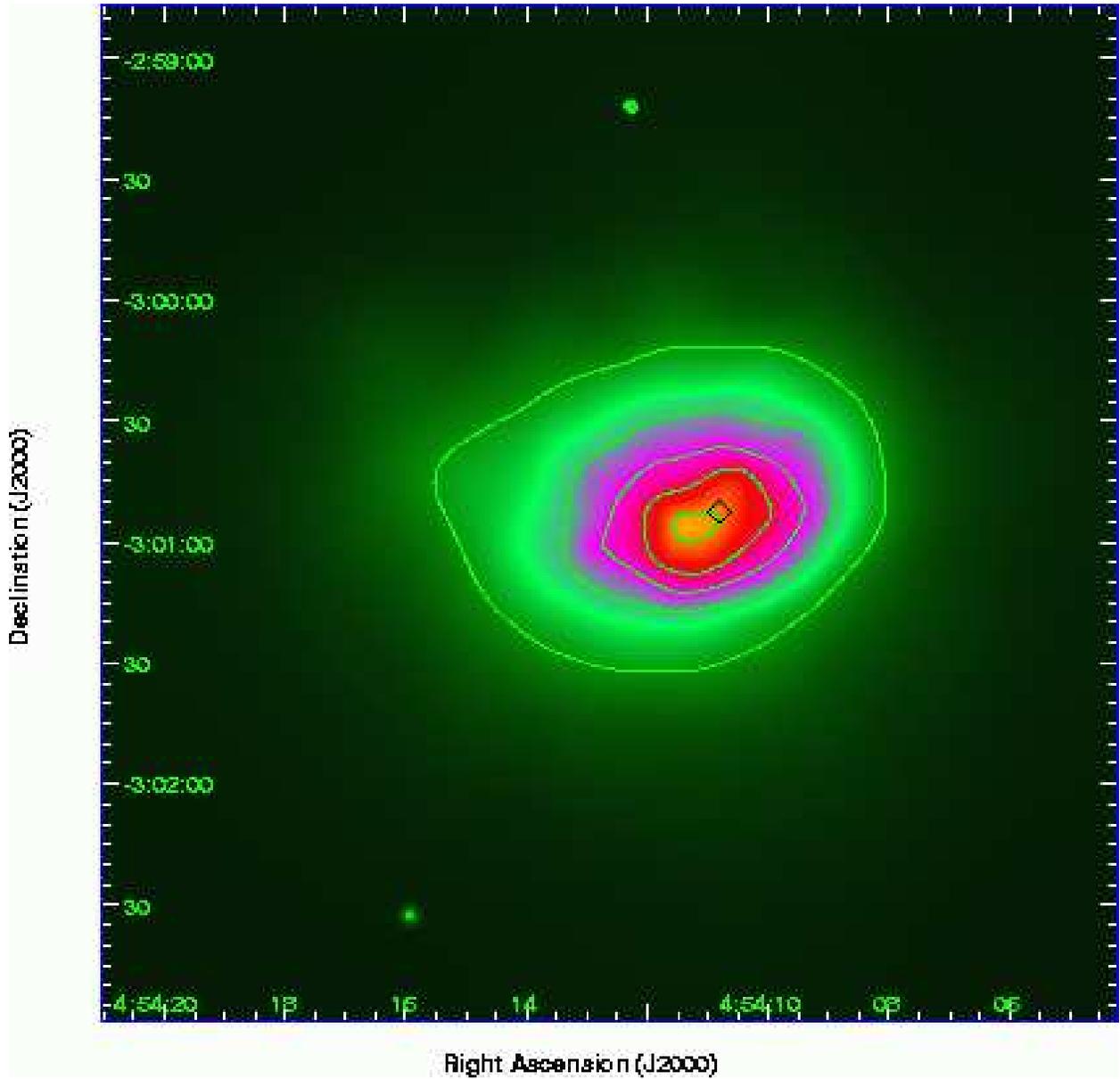}
\caption{A linearly scaled, $3-\sigma$ adaptively smoothed 
image of the diffuse X-ray emission (0.7--7.0 keV) MS0451.6-0305. 
The similarly smoothed 
exposure map has been divided out of the data. The HST (astrometrically
corrected) location of the
brightest cluster galaxy (BCG) is marked with a diamond. Two X-ray point 
sources are visible to the north and the south-east of the cluster. Their
extent show the excellent compact point spread function of the Chandra
Observatory. The peak of the broadband emission lies southeast of the BCG.
The yellow (brightest) surface brightness on the map is about
$4 \times 10^{-8}$ cts s$^{-1}$ cm$^{-2}$ pixel$^{-1}$. The faintest emission
visible in the image is approximately 10 times fainter. \label{ms0451_adapt} }
\end{figure}

An adaptively smoothed image of the diffuse emission of this cluster
was created by using CIAO's 
{\em dmfilth} to replace point sources detected by
{\em wavdetect} with a Poisson distribution based on the sky background
local to the sources. The resulting diffuse image was smoothed to  
a minimum significance level of $3\sigma$; the $1\times1$ binned exposure
map described above was smoothed using the same scales as derived
for the emission image, then it was divided from the smoothed image
(Figure~\ref{ms0451_adapt}).

\begin{figure}[ht]
\includegraphics[angle=0,width=\textwidth]{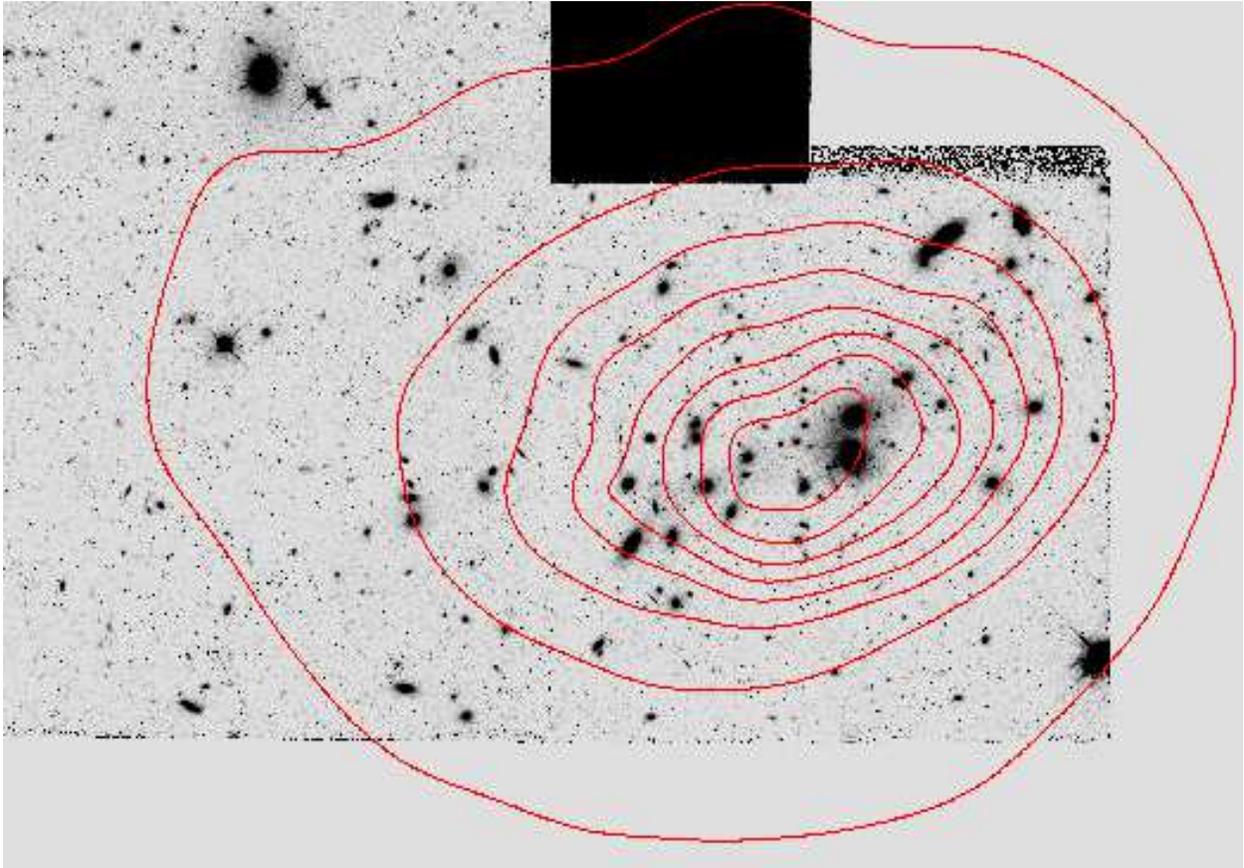}
\caption{A drizzled HST WFPC2 I-band (F702W) observations consisting
of 4 independent exposures with a total exposure time
of 10,400 seconds. The HST image required a correction to the World
Coordinate System keywords in the header based on a correlation between
stars and galaxies in the HST image and GSC2.2 
objects in the field. The shift was $-$0.23 seconds in RA and +1.5 
arcseconds in declination. 
The X-ray contours are overlaid, linearly spaced
from $6 \times 10^{-9}$ to $4 \times 10^{-8}$ 
counts~s$^{-1}$~cm$^{-2}$~pixel$^{-1}$. The brightest cluster galaxy
is NW of the cluster contours centroid. The bright galaxy just south
of the BCG is a foreground spiral. \label{ms0451_hst} }
\end{figure}

We compared the locations of X-ray features to those in the optical
by constructing a Hubble Space Telescope (HST) I-band image of the cluster
(Figure~\ref{ms0451_hst}).
We drizzled together HST WFPC2 I-band (F702W) observations consisting
of 4 independent exposures with a total exposure time
of 10,400 seconds. The HST image required a correction to the World
Coordinate System keywords in the header based on the location of GSC2.2 
objects in the field. The shift was -0.23 seconds in RA and +1.5 
arcseconds in declination. 

We note that the nominal center of the X-ray isophotal 
contours is not located on the brightest cluster galaxy (BCG), 
identified by a diamond on Figure~\ref{ms0451_adapt}. The
large galaxy to the south of the BCG is a foreground galaxy (John Stocke,
private communication). 
The isophotes near the BCG seem to be distorted, perhaps suggesting that 
the BCG or a system centered on the BCG may be contributing to the
surface brightness there. The elliptical BCG  
is at 04h 54m 10.8s, $-03\degr ~ 00\arcmin~ 52.4\arcsec$ 
($x=4164$, $y=3856$
in the physical coordinates of the original Chandra image).

\begin{figure}[ht]
\includegraphics[angle=0,width=\textwidth]{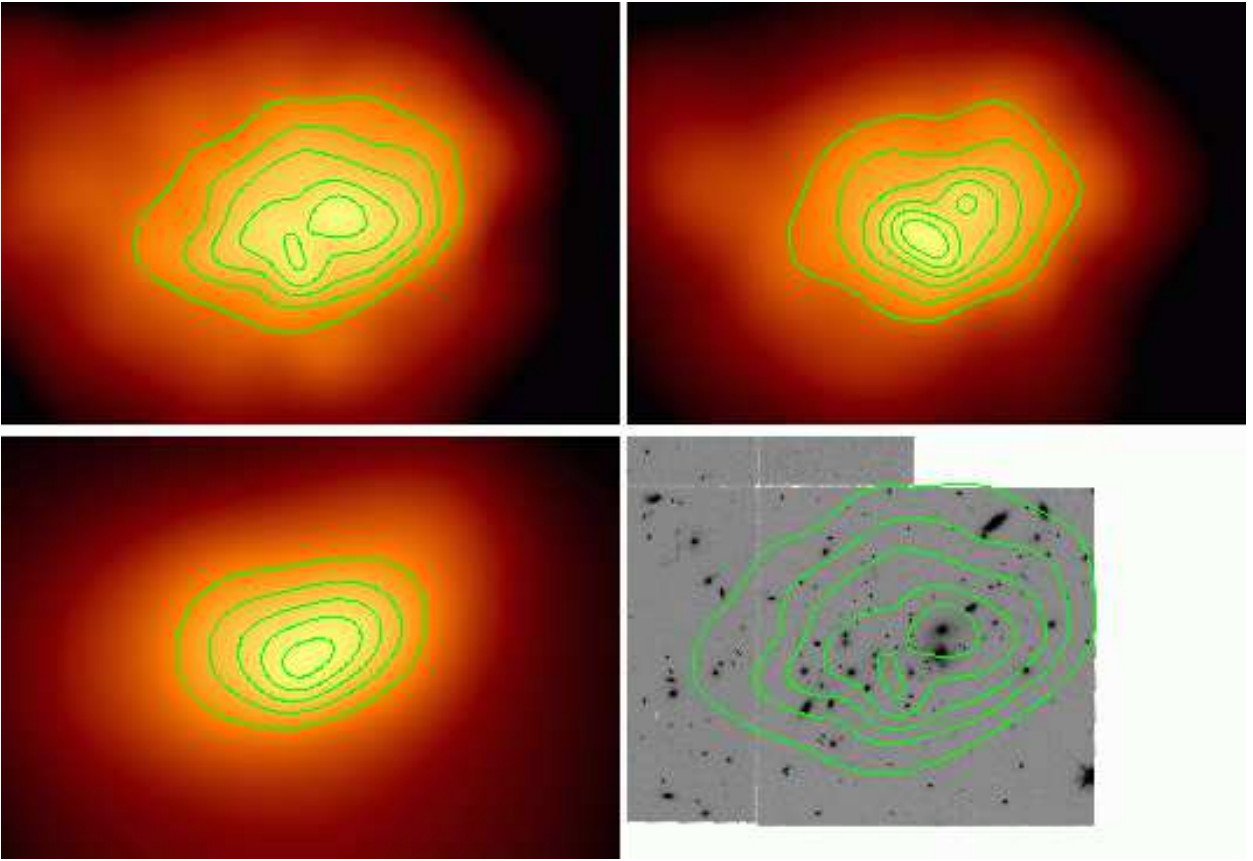}
\caption[]{The central $210\arcsec \times 140\arcsec$ of the cluster. Each image is to
the same scale. The upper left image is the soft image
with the same contours. The upper right image is the ``medium''
band image (1.5-4.5 keV) and the lower left image is the hard
band image (4.5-7.0 keV). The contour levels for each 
image are, in units of photons cm$^{-2}$ s$^{-1}$ pixel$^{-1}$, 
soft (0.18, 0.27, 0.37, 0.48, 0.60, 0.73), 
medium (0.11, 0.16, 0.22, 0.28, 0.35, 0.42, 0.49), 
hard (0.019, 0.028, 0.037, 0.046, 0.055). Every distinct feature in these
contours has $>3\sigma$ significance; however, the contours were chosen
to represent roughly uniform surface brightness spacing between
the peak surface brightness out to the largest scale of the HST image.
One pixel is one $0\farcs5 \times 0\farcs5$ pixel ($0.492\arcsec$).
The lower right image is the astrometrically-corrected (see text) 
HST WFPC2 observation with contours of the soft (0.2--1.5
keV) band image overlaid. The X-ray contours of the soft 
band are centered on the brightest cluster galaxy. 
\label{soft_med_hard} }
\end{figure}

We did not detect any evidence for statistical temperature variations in 
our spectral data. However, 
a qualitative assessment of possible temperature or absorption variations 
across the cluster can be made from the inspection of color maps. 
We divided the events data into three energy bands: 0.2--1.5 keV, 1.5--4.5 
keV, and 4.5--7.0 keV. We created monoenergetic exposure maps 
for the mode of the photons' energies in each band, corresponding to 0.9, 1.6, and 4.6 keV respectively. We then
adaptively smoothed each image, omitting the brightest point source
[\#6 from Molnar et~al.\ 2002]; the identical scales were used to smooth each exposure map.
After dividing the data by the exposure map (setting pixel values
with exposure time less than 1.5\% the total time to 0.0), we then
created contour maps (Figure~\ref{soft_med_hard}).

The main conclusion from Figure~\ref{soft_med_hard} is that 
the emission peaks shift with bandpass; in addition, the centroid of
the emission shifts with bandpass. The positions of the peaks
were identified on the maps. We also computed  centroid positions 
based on the unsmoothed data, weighted by the exposure maps, inside 
a region $256\arcsec \times 256\arcsec$. 
In the soft bandpass, the highest peak lies on the 
BCG, with a secondary peak to the east-south-east. The centroid
is somewhat west of the primary peak, near 
4h 54m 11.26s, $-03\degr \,  00\arcmin \, 56\arcsec$.
In the mid-energy 
bandpass, the brightest peak is closer to the soft bandpass secondary peak,
and a fainter peak is associated with the BCG. The centroid
is near 
4h 54m 11.32s, $-03\degr \,  00\arcmin \, 56\arcsec$, $6\arcsec$ west of the
centroid in the soft band.
In the hardest energy, 
there is no obvious peak near the BCG, and the centroid of this
single-peaked distribution 
is close to the position of the primary peak in the mid-energy
bandpass. The hard peak 4h 54m 11.48s, $-03\degr \,  00\arcmin \, 55\arcsec$.
The soft X-ray emission near the BCG is nearly circular, embedded
in a large-scale elliptical emission closer to that of the global
cluster. The mid-energy peak has a position angle
of $65-70\degr$ E of N 
that is significantly different from the globally-fit (see the next section) position angle
of $100\degr$ E of N. The global position angle of the 0.7--7.0 keV
data is closer to the position angle 
of the hardest emission, which is smooth, and the position angle of 
the $>30\arcsec$ contours of the softest emission. In summary, there seems
to be two peaks to the emission, with a soft peak that lies on the BCG and
a harder peak that is offset from the BCG.

\subsection{X-ray Surface Brightness Models}

\subsubsection{Beta-Model Fits}

We fit the X-ray suface brightness data to spherical and elliptical $\beta$-models,
using SHERPA, the model-fitting engine of CIAO. 
The central portion of a 0.7-7.0 keV image was binned by a factor of 8
(1 original pixel $= 0\farcs4920$), then 
an exposure map of the same size was binned to the same resolution.
The reprojected background map was filtered and binned identically. We found
little advantage to using the background map in this case - results were similar
with and without the background map for the fits without exposure corrections.
Since CIAO 2.3 does not simultaneous allow for the use of exposure maps
and external background, we report fits without the background maps.
In the following section, we report fits with and without the 
exposure correction in order to demonstrate the effect of the
exposure correction. 
Background counts were not subtracted but 
fit as a constant flat contribution to the total signal. Even binned, the total
counts inside each binned pixel was not high, so we experimented with the use
of various statistics: the low-count modification to $\chi^2$ by Gehrels (1986), 
the iterative $\chi^2$ method from Kearns, Primini, \& Alexander (1995), and
Cash maximum likelihood statistics (Cash 1979). We found that the Gehrels (1986)
statistics could result in unusual and unstable results which were very
sensitive to the region being fit, while the Cash (1979) statistics and the 
``Primini'' $\chi^2$ method (Kearns et~al.\ 1995) produced similar results to
each other and consistent uncertainties. We therefore 
use a
simplex method with 
Cash statistics for our results.  
The fit was limited to the central $100\arcsec$ of the
cluster.  We note that
this radius, as we shall see, is about two or three times the
core radius and is approximately equivalent to $r_{2000}$, inside
which the mean mass density exceeds the critical density at 
$z=0.538$ by a factor of 2000. This radius is contained in the range of
$r_{2500}-r_{200}$, within 
which the measured gas fraction is expected to be close to that of the cluster 
as a whole (Evrard, Metzler, \& Navarro 1996).

We report the results of fits 
to a spherical $\beta-$model with $S_X = S_{X0} [1 + (r/r_{core})^2 ]^{-3\beta+1/2}$ and
$r = \sqrt{ (x-x_0)^2 + (y-y_0)^2 }$  
and to an elliptical $\beta-$model (Equation 1) in Table~\ref{tab:sherpa}.
\begin{eqnarray}
S_X(x,y) &=& S_X(r) = S_{X0} (1 + (r/r_{core})^2)^{-3\beta + 1/2} \\
r(x,y) &=& \frac{\sqrt{x^2_{n} (1 - \epsilon)^2 + y^2_{n}}}
{1 - \epsilon} \nonumber \\
x_n &=& (x-x_0) \cos \theta + (y-y_0) \sin \theta \nonumber \\
y_n &=& (y-y_0) \cos \theta + (x-x_0) \sin \theta \nonumber
\end{eqnarray}
The fit parameters include core radius $r_{core}$, $\beta$, cluster central position
$x_0$ and $y_0$, ellipticity $\epsilon$, position angle $\theta$,
and amplitude $S_{X0}$. We report best-fit 
parameters to 90\% projected
confidence for a single interesting parameter, (or $\Delta \chi^2=2.7$; $1.6\sigma$), where 
all other parameters are allowed to take on their best fit values. 

The best-fit spherical model  results in  $r_{core} = 31 \arcsec \pm 3.5 \arcsec$,
$\beta = 0.70 \pm 0.07$ (the uncertainties of $r_{core}$ and $\beta$
are correlated), and a central amplitude 
of $S_{X0}=(1.55\pm0.06) \times 10^{-7}$ 
in units of photons s$^{-1}$ cm$^{-2}$ arcsec$^{-2}$ ($0.7-7.0$ keV).
If the data were not  exposure-corrected, we
obtained somewhat different best-fit parameters for the shape of
the spherical $\beta-$model:  $\beta =0.67 \pm 0.07$, 
$r_{core} = 28\arcsec \pm 3.3\arcsec$. The fit parameters did not change
within the statistical uncertainties with the use of the exposure map.

\begin{figure}
\plotone{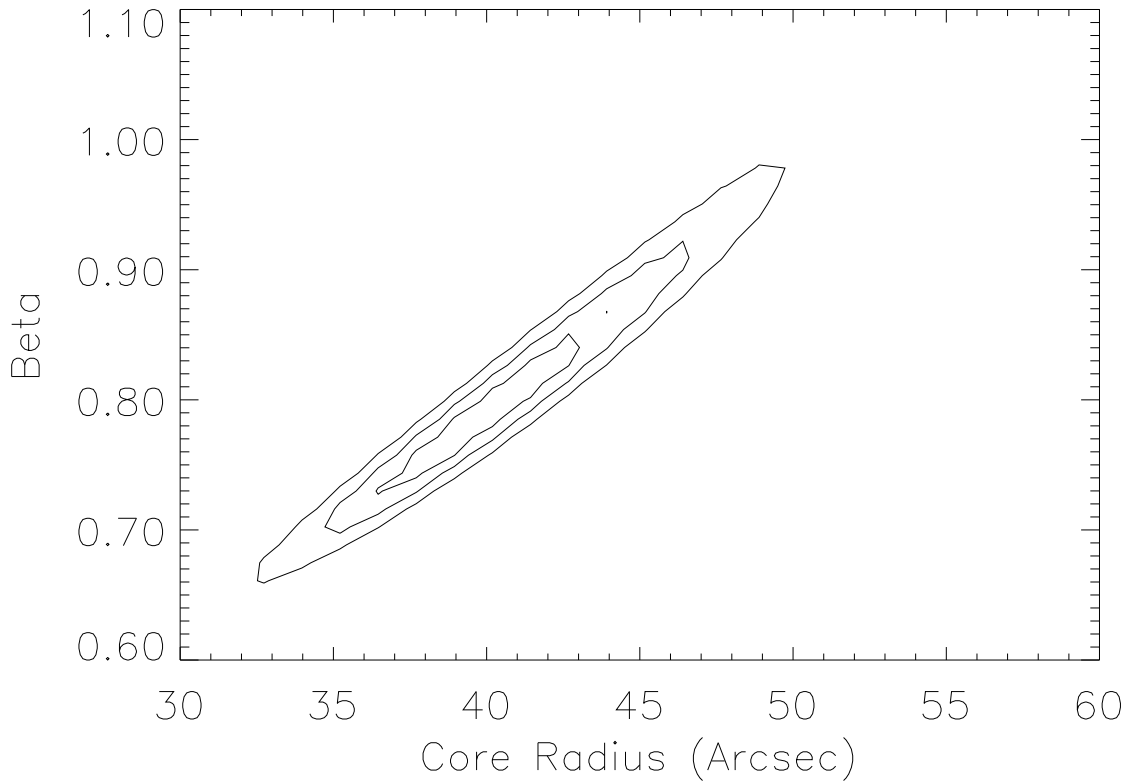}
\caption{\label{beta_rcore_corr} The 1, 2, and 3-sigma confidence contours for
the core radius in arcseconds  and 
$\beta$, for a fit the exposure corrected 0.7--7.0 keV data, binned in 
$8 \times 8$ instrument pixels  
to an elliptical $\beta-$model. The best-fit 
core radii and the power law indices are correlated. The best-fit, plotted as a solid
line here,
is $\beta=0.86$ and $r_{core} = 42\farcs2$ along the semi-major axis.
}
\end{figure}

If we allow the ellipticity and the position angle of an 
elliptical $\beta-$model
to be free, the best-fit $r_{core}$ is $40.2\arcsec \pm 4.1\arcsec$ along the semi-major
axis and 
$\beta$ is $0.79 \pm 0.08$ (Figure~\ref{beta_rcore_corr}). 
The cluster emission is elliptical,
with an ellipticity  
$\epsilon_{SHERPA} = 0.271 \pm 0.016$, and a position angle of 
$\theta = 0.17 \pm 0.03$ radians S of E\footnote{The SHERPA
parameter for position angle $\theta$ seems to be 
incorrectly defined in the documentation -- perhaps the definition
refers to the semi-minor axis not the semi-major axis.} 
or $100\pm2$ degrees E of N, aligned almost directly East-West. 
The amplitude, in units of photons s$^{-1}$ cm$^{-2}$ arcsec$^{-2}$, 
 is $(1.52 \pm 0.06) \times 10^{-7}$ ($0.7-7.0$ keV)
The background 
is $(2.09 \pm 0.65) \times 10^{-9}$ ($0.7-7.0$ keV).
 We also report the best-fit to data without an exposure correction
 in Table~\ref{tab:sherpa}. We note that the exposure-corrected 
 data fit to an elliptical model had the most symmetric uncertainties in
 the background level.

When the region around the BCG was excluded from the fit, the results 
did not change, except for the larger error bars on $r_{core}$ and $\beta$. 
Therefore, a large fraction of the emission from the cluster can be modelled
robustly by an ellipsoidal $\beta-$model.

The center of a 2-D  $\beta-$model fit to the 0.7-7.0 keV data was 
independent of whether an elliptical or a spherical model was 
fit: 04h 54m 10.9s, $-03\degr \, 00\arcmin \, 47\farcs1$. 
Relative to the surface brightness data contours, the center of the best fit
 lies between the BCG elliptical and the centroid of the bulk of the
X-ray emission.  This location is about $7.5\arcsec$
west of the BCG, whose position is defined in our corrected HST data.

The brightness of a point source that could hide in any
$8\times8$ ($4\arcsec \times 4\arcsec$) 
bin depends on the number of counts in each bin. The 
brightest point source that could hide in these data would be, conservatively, 
a $5\sigma$ source in the center of the cluster. Such a source would
need have to have about 61 total counts, or  
$>1.5 \times 10^{-3}$ counts s$^{-1}$ (0.7-7.0 keV), or
$F_x \sim 1.1 \times 10^{-14} \flux$ in the same bandpass to be detected 
above the cluster. A $3\sigma$ source in the highest flux bins would
have to have 33 total counts, or $F_x \sim 6 \times 10^{-15} \flux$ (0.7-7.0
keV). The largest positive excess over an elliptical beta-model in our data is
$<2.8\sigma$. If the data are binned to a finer grid of $4\times4$ pixels, 
the maximum hidden point source flux is reduced to approximately
$2/3$ the fluxes of the $8\times8$ pixel regions. 

\subsubsection{Goodness of Fits}

We examined maps of the residuals between the best-fit models and the 
$8\times8$ binned data. The number of bins where the 
spherical model residuals which are greater
than $2\sigma$ from the model is 49, whereas the corresponding number
of bins for the elliptical model residuals is 23. No bins exhibit a departure
of greater than $\sim3\sigma$ from the model. The most prominent
features in the maps of residuals align with a linear feature with
a depth of about 20\% of the total in the 
exposure map. The alignment suggests that at least some of the residuals are
related to an imperfect exposure correction. 

\begin{figure}
\plotone{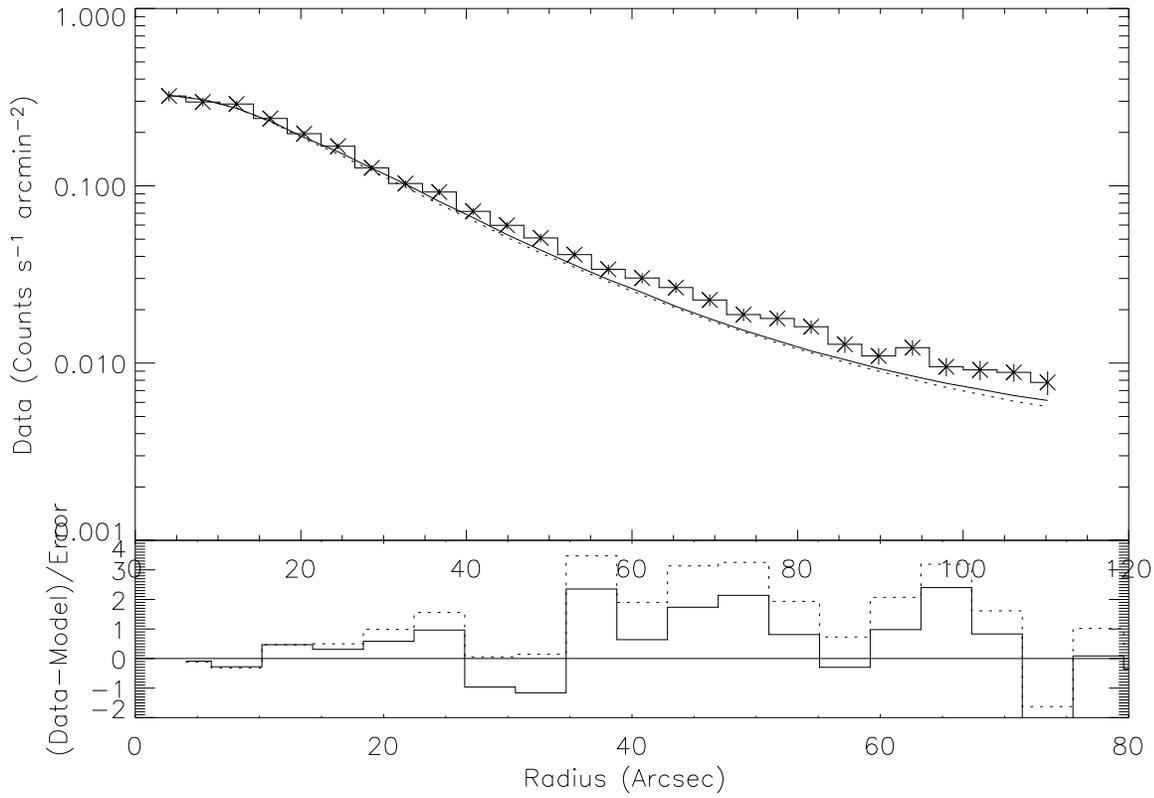}
\caption[]{A radial plot extracted identically from the data (histogram, with
$1-\sigma$ error bars), the best-fit elliptical $\beta-$model (solid line), and
the best-fit spherical $\beta-$model (dotted line). The residuals for the
elliptical fit (solid line) and the spherical fit (dashed line) are plotted
in units of $\sigma$ below the plot. \label{chisq}}
\end{figure}

If we bin the data and the simulations produced from the best-fit
models into  radial profiles, then compute a $\chi^2$, the resulting
fits are both rejected.
The spherical $\beta-$model results in a $\chi^2 = 162.5$ for 22 degrees of freedom.
By adding two additional parameters
with the elliptical $\beta-$model, we obtain a $\chi^2 = 116.8$ for 20 degrees of freedom.
This improvement is statistically significant, but the fit is
still formally rejected. Figure~\ref{chisq} shows the radial plot, with 27 points
for the data in a histogram. The elliptical model is plotted with a solid line
and the spherical $\beta-$model is plotted with a dashed line. The residuals are
plotted below the radial surface brightness plot. The spherical model shows
higher residuals at $r \sim 40-80\arcsec$.

In summary, the data clearly exclude a traditional spherical $\beta-$model. An elliptical
beta model better represents the surface brightness distribution, both
in a statistical and in a qualitative sense, but it
too is an incomplete description of the data. The 2D binned 
data do not show any statistically significant departures from the 
model, mainly because of the limitations of bins with small numbers
of counts in them; in contrast, radially binned data seem to exclude the models. 
We are left, then, to use the $\beta-$model approximation to derive
gas masses and isothermal gravitational masses, but to retain the caveat
that we are, as yet, limited by an incomplete description of the data.

\section{Masses and Mass Profiles}
Many workers in this field have simply approximated the surface brightness profiles of clusters where deprojection is not possible by
extracting radial profiles. These radial profiles are generally fit to
beta-model functional forms. 
We take an additional step (see also Hughes \& Birkinshaw 1998) to previous studies
in the next sections. In addition to reporting the results for MS0451.6-0302 from
a spherically symmetric analysis, we also estimate the
impact of a first-order correction to the spherically symmetric case by
applying the results of an elliptical beta-model fit to the derivation
of the gas mass and other cluster properties. We find that the derived gas mass
and gas fraction are affected somewhat by moving from the spherically
symmetric to the ellipsoidal case. In contrast, as other workers have also
found in a study of ROSAT clusters (Piffaretti, Jetzer, \& Schindler 2002), 
the gravitational mass enclosed inside a sphere is not significantly
affected by the spheroidal assumption.  

In the following section, we describe the formalism we use to derive the gas mass profile for an
isothermal spherical and ellipsoidal beta-model for the gas distribution, including,
for completeness, a discussion of the Sunyaev-Zel'dovich temperature decrement. 

\subsection{Spherical Model}
For a spherical isothermal $\beta-$model, the electron number density
$n_{e}$ follows the model 
\begin{equation}
n_{e}(\theta)=n_{e0}\left(1+\frac{\theta^{2}}{\theta_{c}^{2}}\right)^{-3\beta/2},
\end{equation}
where n$_{e0}$ is the central electron density. 
The central electron density can be derived from
\begin{equation}
n_{e0} = \sqrt{ \frac{S_{X0}}{A_1 D_A} }
\label{circ_ne}
\end{equation}
where $S_{X0}$ is defined in units of counts s$^{-1}$ sr$^{-1}$, 
$D_A$ is the cluster angular distance in units distance per radian 
and $A_1$ is defined below.

The X-ray surface brightness, $S_{X}$, in terms of the 
angular distance diameter, $D_{A}$, is

\begin{equation}
S_{X}(\theta)={\frac{1}{4\pi(1 + z)^{3}}}\int{D_{A}n_{e}^{2}\Lambda_{e}d\zeta}
=S_{Xo}\left(1+\frac{\theta^{2}}{\theta_{c}^{2}}\right)^{1/2-3\beta},
\end{equation}
where $S_{X0}$ is the central X-ray surface brightness, $z$ is the cluster redshift, and $\Lambda_{e}$ 
is the X-ray cooling function of the ICM in the cluster rest frame.  
$\Lambda_{e}=\epsilon/n_{e}^{2}$,
where $\epsilon$ is the spectral emissivity 
(Birkinshaw, Hughes, \& Arnaud 1991;  Hughes  \&  Birkinshaw 1998).

The Sunyaev Zel'dovich Effect, SZE, is the change in the observed brightness 
temperature of the Cosmic Microwave Background (CMB) radiation resulting from passage
of the CMB radiation through the ionized gas permeating a galaxy cluster 
(Sunyaev  \&  Zel'dovich 1972; Sunyaev  \&  Zel'dovich 1980).  The temperature decrement can 
be described by a spherical isothermal $\beta$ model if the density of the 
cluster follows the form $[1 + (r/r_{core})^2]^{-3\beta/2}$.
(Cavaliere  \&  Fusco-Femiano 1976; Birkinshaw 1999). The resulting temperature
decrement is given by

\begin{equation}
\Delta T=f(x)T_{CMB}D_{A}\frac{\sigma_{T}k_{B}}{m_{e}c^{2}}\int{n_{e}T_{e}d\zeta}
=\Delta T_{0}[1 + \frac{\theta^{2}}{\theta_{c}^{2}}]^{1/2-3\beta/2},
\end{equation}
where $f(x)$ is the frequency dependence of the SZE, 
$T_{CMB}$ is the temperature of the CMB radiation,
$\sigma_{T}$ is the Thomson cross section,
$k_{B}$ is the Boltzmann constant,
$m_{e}$ is the electron mass,
$n_{e}$ is the electron density,  
$T_{e}$ is the cluster temperature,
$\Delta T_{o}$ is the central temperature decrement, 
$\theta$ is the angular radius in the plane of the sky,
$\theta_{c}$ is the angular core radius, and $\beta$ is a shape parameter 
that describes the radial falloff of the gas distribution from the beta-model. 
The integration is along the line of sight $\ell$ = D$_{A}$$\zeta$ 
(Hughes  \&  Birkinshaw 1998; Carlstrom   et~al. 2000).

In principle, the cluster distance, D$_{A}$, can be estimated independently of $H_0$ 
by combining  X-ray observations
with Sunyaev Zel'dovich observations.  
Solving equations (3, 5, 6) for the angular diameter distance, $D_{A}$, in terms
of known parameters (Myers   et~al. 1997; Hughes  \&  Birkinshaw 1998; 
Birkinshaw 1999; Patel   et~al. 2000; Reese   et~al. 2002), gives 

\begin{equation}
D_{A}=\Bigl(\frac{\Delta T_{0}}{A_{2}}\Bigr)^{2}\frac{A_{1}}{S_{Xo}},
\end{equation}

where

\begin{equation}
A_{1}=\frac{1}{4\pi (1+z)^{3}}\frac{\epsilon_{o}}{n_{eo}^{2}}\sqrt{\pi}\frac{\Gamma(3\beta-1/2)}{\Gamma(3\beta)}\theta_{c},
\end{equation}

\begin{equation}
A_{2}=\delta_{SZE}(x,T_{e})T_{CMB}\frac{k_{B}T_{e}}{m_{e}c^{2}}\sqrt{\pi}\frac{\Gamma(3\beta/2-1/2)}{\Gamma(3\beta/2)}\theta_{c}.
\end{equation}
where $\delta_{SZE} = -1.879$ is the relativistic correction to the frequency dependence
for $kT = 10.8$ keV  (Itoh, Kohyama \& Nozawa 1998).

The gas mass derived from the assumption of a $\beta-$model and spherical symmetry is simply
\begin{equation}
M_{gas}(<r) = 4 \pi \mu m_H n_{e0} \int^{r}_{0} [1.0 + (r/r_{core})^2]^{-3\beta/2} r^2 dr.
\end{equation} 

The corresponding gravitating mass (Evrard et~al.\ 1996),
 from the assumptions of hydrostatic equilibrium ($\nabla P = \rho_g \nabla \phi$ 
and $\nabla^2 \phi = -4 \pi G \rho_M$) and isothermality ($\rho = -\frac{kT}{4 \pi G \mu_h m_h} \nabla^2 \ln{\rho_g}$), is
\begin{equation}
M(<r) = \frac{3 \beta kTr}{G \mu_H m_H} \left(\frac{r}{r_{core}}\right)^2 \left/\left[1 + \left(\frac{r}{r_{core}}\right)^2\right]\right.
\end{equation}
where $\beta$ and $r_{core}$ are from the fit to a spherical $\beta-$model, $kT$ the X-ray
temperature, $G$ the gravitational constant, $\mu_H$ the mean mass per particle 
($\mu_H=0.59$), and $m_H$ the mass of a proton. Note that for large $r$, an isothermal
mass distribution diverges like $r$.

A somewhat more physical assumption may be that the temperature is not constant
(at least outside the radius where we have direct 
measurements). We can allow for an 
unobserved temperature gradient by approximating the system with 
a ``polytropic'' equation of state (Lea 1975) where
\begin{equation}
P=K_{\rm eff} \rho_g^{\gamma_{\rm eff}}
\end{equation}
and 
\begin{equation}
T = \frac{P}{\rho}\mu_H m_H = K_{\rm eff} \mu_H m_H \rho_g^{\gamma_{\rm eff}-1}
\end{equation}
In the spherical case where the gas is distributed like the beta-model,
\begin{equation}
M(<r) = \frac{3 \beta \gamma kT(0) r}{G \mu_H m_H} \left(\frac{r}{r_{core}}\right)^2\left/\left[1 + \left(\frac{r}{r_{core}}\right)^2\right]^b\right.
\end{equation}
where $\gamma$ is the effective polytropic index, $K_{\rm eff}$ is the
effective polytropic constant $K$, $kT(0)$ is the X-ray temperature at 
radius $r=0$, and $b=1.5 \beta (\gamma-1)+1$ (See also Henriksen \& Mushotzky 1986). 
Nonphysical masses result if $\gamma > 1 + 1./3\beta$ (or $\gamma > 1.41$ for $\beta=0.8$).
Typical polytropic indices found for nearby relaxed 
clusters are around $1.1-1.2$ (e.g.\ De Grandi \& Molendi 2002; Markevitch
et~al.\ 1999). The ratio of the polytropic mass to an isothermal mass
is $\gamma  [ 1 + (r/r_c)^2]^{-1.5 \beta (\gamma-1)}$.
If MS0451 has an effective polytropic index of $\gamma=1.2$ and
surface brightness index $\beta=0.8$, then 
the enclosed gravitational mass at 5 core radii would be 
about a factor of $0.54$ that of the isothermal mass. If $\gamma=1.1$, the
factor is $0.74$. The mass at $r_{500}$ is thus rather sensitive to 
$\gamma$.

\subsection{Ellipsoidal Model} 

If we relax the assumption that the cluster electron density distribution is spherically
symmetric and assume that the distribution is a prolate or oblate spheroid, we can derive
equivalent expressions for $n_{e0}$ and $D_A$. Here, we assume that the
gas follows an ellipsoidal beta-model distribution as follows:
\begin{equation}
\rho_g = \rho_o [ 1 + \frac{\sum_{i=1}^{3} e_i^2 x_i^2}{r_c^2} ]^{-3\beta/2},
\end{equation}
where $\rho$ is the gas density, $x_i$ the intrinsic coordinate distances in each of
three dimensions, $r_c$ the core radius in the direction of the
semi-major axis, and $e_i$ the ratio of the major to minor axes in
each direction. For a prolate model of intrinsic eccentricty
$e$, where the axis of symmetry is the 3rd coordinate, 
$e_i = [e, e, 1.0]$ and for an oblate model, $e_i = [1, 1, e]$. 

We follow the derivations from 
Hughes \& Birkinshaw (1998; HB98) and Fabricant, Rybicki, \& Gorenstein (1984), but 
present them here in order to use terminology similar to
 the previous section. The expressions for
$D_A$ and $n_{e0}$ are identical. The expressions $A_1$ and $A_2$ can be modified for
an oblate
distribution where $i$ is the angle of the axis of symmetry with respect to the line of 
sight ($i=90\degr$ corresponds to the axis of symmetry lying in the plane of the sky):
\begin{eqnarray}
A_{1,oblate} = A_1 \frac {\sqrt{1-e_{HB}^2 \cos{i}^2 }} {\sin{i} } \\
A_{2,oblate} = A_2 \frac {\sqrt{1-e_{HB}^2 \cos{i}^2 }} {\sin{i} }
\end{eqnarray}
and a prolate distribution gives:
\begin{eqnarray}
A_{1,prolate} = A_1 \frac  {\sqrt{e_{HB}^2-\cos{i}^2 }}{e_{HB}^2 \sin{i} }  \\
A_{2,prolate} = A_2 \frac  {\sqrt{e_{HB}^2-\cos{i}^2 }}{e_{HB}^2 \sin{i} }
\end{eqnarray}
where $e_{HB}$ is the {\em observed} semi-major/semi-minor axis ratio. The 
eccentricity $\epsilon$ from the SHERPA elliptical $\beta-$model is related
to $e_{HB}$ by $e_{HB} = 1.0/(1.0 - \epsilon)$. The intrinsic axis ratio
$R$ from Fabricant et~al.\ (1984) can be recovered from the expression
$e_{int} = Re_{HB}$ where $R_{oblate} = \frac {\sqrt{1-e_{HB}^2 \cos{i}^2 }} {\sin{i} }$
and $R_{prolate} = \frac  {\sqrt{e_{HB}^2-\cos{i}^2 }}{e_{HB} \sin{i} }$.

The intrinsic coordinate is related to the observed coordinate by the
usual rotation transformation where the rotation is around the first
axis:
\begin{eqnarray}
x_{1,i} = x_{1,o} \\
x_{2,i} = x_{3,o} \sin{i} + x_{2,o} \cos{i} \\
x_{3,i} = x_{3,o} \sin{i} - x_{2,o} \cos{i}
\end{eqnarray}

We note that the expressions for $A_1$ and $A_2$ 
diverge for certain values of the inclination
angle $i$. At $i=0$, a prolate and oblate distributions would appear as perfectly circular
clusters to the observer, and no information could be recovered regarding the elongation
along the line of sight. Also, for $i \leq \arccos(1.0/e_{HB})$, an oblate distribution cannot
reproduce the observed axis ratio $e_{HB}$. Therefore, for an oblate distribution to
be plausible, there is a minimum inclination angle $i$.

The gravitating mass density $\rho_M$ can be computed directly from the X-ray temperature
and gas distribution, using the assumptions of hydrostatic equilibrium and isothermality as
above. The gas distribution can be recast as:  
\begin{equation}
\rho_g = \mu_e m_h n_{e0} u^{-3\beta/2}
\end{equation}
where $u = [1 + \Sigma_{i=1}^{3}  \frac{e_i^2 x_i^2}{r_c^2}]$ and
$\mu_e=1.4/1.2$. Here $r_c$ is the core radius along the
longest axis, and, for convention, the axis of symmetry is along the $z$ or $3$ axis and $e_i$
is the axis ratio. 
So for the general case where 
\begin{equation}
\rho_M = - \frac{kT}{\mu_H m_H G 4 \pi} \nabla^2 \ln \rho_g
\end{equation}
the result can be written analytically where
\begin{equation}
\rho_M = - \frac{3 \beta kT}{\mu_H m_H G 4 \pi  r_c^2 u^2} \left[ - \Sigma_{i=1}^{3} \left(\frac{2e_i^2x_i}{r_c}\right)^2
+u \Sigma_{i=1}^{3} {2 e_i^2} \right].
\label{equation:rho_dm}
\end{equation}
For the case $e=[1, \, 1, \, 1]$, Equation~\ref{equation:rho_dm} reduces to the 
spherical case.

This derivation assumes that the dark matter is distributed in concentric, similar 
ellipsoids. The isothermal 
ellipsoidal $\beta-$model derived here is not physically realistic over all scales, since large 
eccentricities can result in negative, unphysical dark matter densities. For
small eccentricities, this derivation is a perturbation of the spherical $\beta-$model, and is
not a bad approximation to the effects of moderate elongation. However, 
we warn that this derivation is only intended to explore 
the implications of an ellipsoidal model.

\section{Cluster Luminosity}

We derive the cluster luminosity inside a radius of $1 h^{-1}$ Mpc. The
total count rates (0.7-7.0 keV) are derived by
integrating the best fit elliptical $\beta-$model inside a circle of
$244\arcsec$ (flat, $\Omega_M=0.3$ cosmology). The aperture correction
from $r=85\arcsec$ is 1.074, for 0.319 counts s$^{-1}$ total.
From the spectral fits with XSPEC, we have derived the conversion
from count rate to luminosity and flux for the best-fit Raymond-Smith model
with $kT=10.6$ keV. Our results here are not too sensitive to the details
of the spectral fit. We report the observed fluxes and the
intrinsic luminosities in Table~\ref{tab:lum}.

If the luminosity-temperature ($L_x - T_x$) relation does not evolve, 
the predicted temperature
for this cluster from the Markevitch (1998) L-T relation is
$10.5-12.2$ keV, depending on whether we use the ROSAT band relation
or the bolometric relation in Markevitch (1998). Therefore the temperature
and luminosity we measure for this cluster at $z=0.538$ is consistent with the
local luminosity temperature relation.

\section{Central Density, Gas Mass, and Total Mass from X-ray Properties}

The estimates for X-ray properties based on the X-ray data alone are reported 
in Table~\ref{xray_masses}. 
Here we use the results of the $\beta-$model fits and the X-ray properties alone to 
derive the cluster gas mass and the baryonic fraction, assuming isothermality, 
and hydrostatic equilibrium as described in the previous section.
We assume a temperature of $kT=10.6$ keV (this work), along with an 
emissivity of $1.343 \times 10^{-12} \, n_e^2 \, {\rm count \, s}^{-1} \,
{\rm cm}^{5}$. For the spherical $\beta-$model,
here we assume a core radius $31\arcsec$, $\beta = 0.75$, central
surface brightness of $0.34$ counts arcmin$^{-2}$ s$^{-1}$, $z=0.5386$.
We derive a central electron
density of $n_{e0} = (0.0146 \pm 0.002) h^{1/2}$ cm$^{-3}$ using 
Equation~\ref{circ_ne}. For elliptical $\beta-$models we use the
best-fit parameters from Table~\ref{tab:sherpa}.
The $r_{500}$ radius, inside which the mean density is 500 times that of the
critical density at that redshift $z$ 
($\rho_c(z) = \rho_c(0) ( \Omega_M (1+z)^3 + \Omega_\Lambda)$ 
if $\Omega_M + \Omega_\Lambda = 1$) for this cluster is
$0.97 \pm 0.13 h^{-1}$ Mpc for 
$M_{500} = (8.6\pm 1.2)  \times 10^{14} h^{-1} M_\odot$. For an ellipsoidal modal,
$M_{500} = (9.1\pm1.2)  \times 10^{14} h^{-1} M_\odot$.
The uncertainties include the
uncertainty in the statistics of the 
spectral and spatial fitting, added in quadrature, but not the systematic uncertainty in
the spectral fit.

The gas mass inside $r_{500}$ is $(5.6 \pm 3.3) \times 10^{13} h^{-5/2} M_\odot$ for these
fit parameters, corresponding to a baryonic fraction in hot gas of 
$0.065 \pm 0.01 h^{-3/2}$. The uncertainties here include systematic and
statistical uncertainties added in quadrature (Patel et~al.\ 2000). 
However, we will show that the uncertainties in the 
shape and orientation of the cluster lead to even larger uncertainties
($\sim 0.03$) in the gas fraction.
If we could constrain the shape and orientation of a given cluster, one
of the largest systematic uncertainties would be reduced and we  
could refine the estimates of the gas fraction. Here we will show that
while current data are not quite good enough to accomplish this goal, there
is promise in the technique for data in the near future.
The estimates for masses, electron density, and gas fraction 
are reported in Table~\ref{xray_masses}. 

As pointed out first by Briel, Henry, \& B\"ohringer (1992) (and
subsequently expanded upon by White et~al.\ 1993 and others), the gaseous
baryonic fraction in clusters can be used to place a limit on $\Omega_m$,
given a constraint on the baryonic density $\Omega_B$ 
from primordial nucleosynthesis and deuterium measurements. If we
use $\Omega_B = 0.019 \pm 0.002 h^{-2}$ from Burles \& Tytler (1998),
we obtain an upper limit to $\Omega_M$ of $0.29 \pm 0.05 h^{1/2}$. 

To get a better census of the baryons in this cluster, we can also 
account  for the baryons associated with the galaxies, which in general
are only a small contribution to the total amount of baryons in clusters. 
The amount of mass associated with the galaxies from
the total optical luminosity of the cluster was estimated by 
assuming a mass to light ratio for the galaxies. Carlberg et~al.\ (1996)
measure a k-corrected r-band (Gunn) luminosity inside their definition 
for $r_{200}=r_v = 1.4\pm0.17 h^{-1}$ Mpc 
of $4.5 \times 10^{12} h^{-2} L_\odot$. For their assumed cosmology
of $q_0=0.1$, $r_v$ is $346\arcsec$. Scaling to the same projected
radius as our assumed $r_{500} = 0.95 h^{-1}$ Mpc, or $214\arcsec$,
$L_{500,R} = 8.7 \times 10^{11} h^{-2} L_\odot$, 
assuming $L(<\theta) \propto \theta$.
The average elliptical galaxy 
$M/L_R = 6.64 h$ (van der Marel 1991) for Johnson R. The conversion from
Gunn r to Johnson R is about a factor of 1.8 for elliptical galaxies
(Frei \& Gunn 1994), including a color term for $g-r\sim0.75$ 
(Carlberg et~al.\ 1996).
Therefore the mass associated
with the galaxies alone  could be as high as $1 \times 10^{13} h^{-1} \Msun$
inside a projected angular distance of $214\arcsec$. 
(We did not take into account a weak dependence 
of $M/L$ on the luminosity of the
individual galaxy (van der Marel 1991).) 
The corresponding $M_{gal}/M_{gas} \approx
0.15 h^{3/2}$, or $0.10$ if $h=0.75$. 
If most of the matter in this $M/L$ ratio is
baryonic, the gas fraction of the cluster can now be corrected to a baryon fraction.
 The estimate for  $\Omega_M$ that would be consistent with primordial
 nucleosynthesis and deuterium constraints  becomes
$ \Omega_M = 0.29 h^{1/2} [1 + 0.15 h^{3/2}]^{-1}$ or, for $h=0.75$,
$\Omega_M = 0.23 (\pm 0.05)$. The estimated value of $\Omega_M$ 
increases somewhat if some of the matter ascribed to elliptical galaxies here 
is not baryonic. 

We can turn this calculation around, if we use WMAP values for
$\Omega_M$, to calculate what fraction of the baryons are in the 
ICM. The WMAP ratio of baryons to total matter is independent of
$H_0$, $\Omega_B/\Omega_M = 0.166$. The gas fraction estimated for
this cluster, for $H_0=71$, is approximately 0.10; we will show later
that this value could be as high as 0.12-0.14, depending on geometric
assumptions. Therefore, at least 60\% $h_{71}^{-3/2}$ of the baryons
in the cluster are in the hot ICM.  Unless cluster-specific processes
like ram-pressure stripping are particularly effective and efficient at removing
baryons from the galaxies, one hypothesis based on this observation
is that most of the baryons in the universe may lie between the 
galaxies. Studies of the Lyman-alpha forest suggest that is the 
case at high redshift, $z\sim3$ 
(Fukugita, Hogan \& Peebles 1998; Hui et~al.\ 2002). Assessment of
gas fractions in clusters suggest it is also true at lower redshifts;
refinements of such measurements could show the dependence on mass scale 
of the efficiency of galaxy formation (David et~al.\ 1990;
Bryan 2000).

\subsection{Comparison to Other Mass Determinations}

We compute a gravitational mass inside $r_{500}$ of $M_{500} = (8.6-8.9) \times 10^{14}
h^{-1} \Msun$ or, alternatively, $M(r<1~h^{-1}$ Mpc$) = (8.8-9.2)\times 10^{14} h^{-1}
\Msun$. In this section, we compare this mass estimate
to other mass estimates for this cluster, from the literature. One challenge
in the literature is to compare masses estimated at radii defined in 
many different ways with several cosmological assumptions and definitions
built in. As an aside, we would like to 
encourage observers to report enclosed masses at either a 
metric radius or a fixed angular radius, in addition to $M_{200}$ or $M_{500}$
if desired, in order to minimize the computations required to make
direct comparisons.

The gravitational mass from
optical, the virial mass, can be 
derived from the 
one-dimensional velocity dispersion $\sigma_v = 1371 \pm 105$ km s$^{-1}$ 
 and the estimated virial radius $R_v = 1.4 h^{-1}$ Mpc (Carlberg et~al.\ 1996). See 
Carlberg et~al.\ (1996) for details on how the virial radius was computed, 
based on CNOC observations for MS0451.6-0302 
(Ellingson et~al.\ 1998).  For this cluster, Carlberg et~al.\ (1996) obtain
an $M_{200} = 1.8 \times 10^{15} h^{-1} \Msun$, for an open $\Omega_m=0.2$ 
cosmology. Adjusting to our flat $\Omega_m=0.3$ cosmology and 
extrapolating to $r_{500}$ by assuming that $\rho \propto r^{-2}$, we obtain
$M_{500} = (1.4 \pm 0.25)  \times 10^{15} h^{-1} \Msun$. The optical $r_{500}$
is larger than the $r_{500}$ estimated from the X-ray temperature and surface
brightness distribution. The optically-derived mass 
at the same angular scale as our X-ray estimate (also taking
into account the differences in the angular distance scale between the two cosmologies)
of $0.97 h^{-1}$ Mpc, is $( 1.3 \pm 0.25 )\times 10^{15} \Msun$. 
The total mass ($M_{500}$) 
calculated from the X-ray values ($(8.6-9.1)\pm1.2 \times 10^{14} \Msun$) is only 
somewhat less than 
that derived from the optical velocity dispersion and
the positions of the member galaxies by Carlberg et~al.\ (1996). Within
the uncertainties of both estimates, however, they are consistent, especially
if the underlying dark matter potential is somewhat elongated, which gives the
higher X-ray mass estimate. We will discuss this dependency further in the 
next section.

\begin{figure}
\plotone{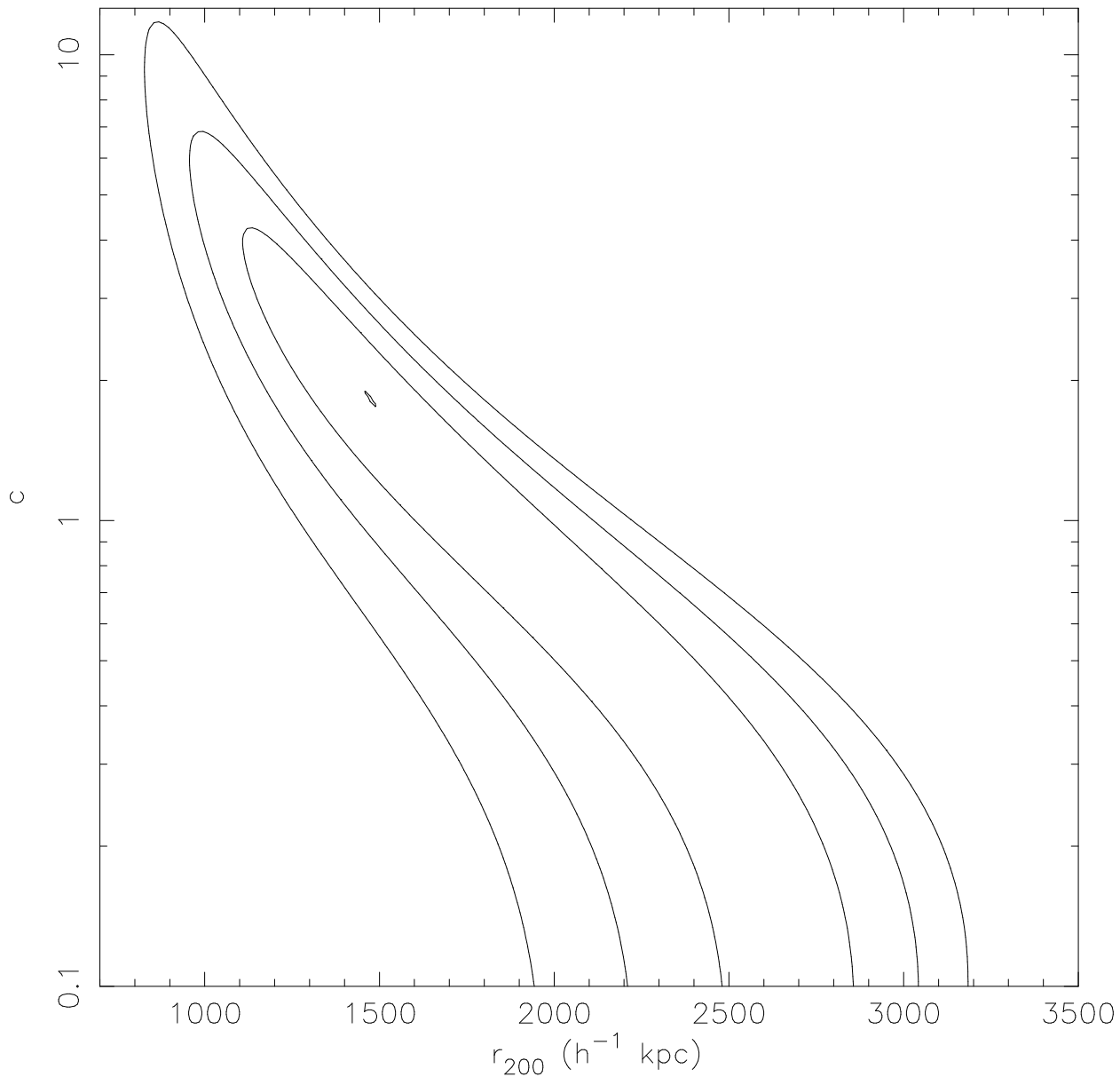}
\caption{The $\chi^2$ values for a fit to the Clowe et al. (2001) weak 
lensing data for MS0451.6-0305 are plotted as contours for the Navarro-Frenk-White
(NFW) profile parameters of concentration index $c$ and $r_{200}$, as in
Figure 11 of Clowe et al. (2000). Each contour represents a change of
$1~\sigma$ in the quality of the fit. 
The best-fit is $c=1.82$ and $r_{200}=1474 h^{-1}$ kpc, for a flat
$\Omega_M=0.3$ cosmology. \label{clowe}}
\end{figure}

Clowe et~al.\ (2000) find a best-fit mass for MS0451.6-0302, corrected for projection, 
to their ground-based weak-lensing data (obtained with the Keck telescope)
with a NFW (Navarro, Frenk, \& White 1996) dark matter 
profile where the concentration index
$c = 1.5$ and $r_{200} = 1060 h^{-1}$ kpc, assuming an underlying
cosmology of $\Omega_M=1$ and $z=0.55$.  
We recomputed the NFW parameters for the same weak lensing data, 
assuming a flat cosmology with
$\Omega_M=0.3$ and a cluster redshift $z=0.5386$, for 
a best-fit of $c=1.82$ and $r_{200}=1474 h^{-1}$ kpc. We also
fit the 1D velocity dispersion of $\sigma=986^{+53}_{-58}$ km~s$^{-1}$ 
(uncertainties are $1 \sigma$) 
assuming an isothermal mass distribution. As in Clowe et~al.\ 
(2000), the concentration is not constrained very well because
of the small field of view for the Keck data. The 
constraints on $r_{200}$ and $c$ with the revised cosmology
are plotted in Figure~\ref{clowe}. The $1\sigma$ minimum for the value of 
$r_{200}$ is  $1100 h^{-1}$ kpc. The mass could be much higher
for lower concentrations $c$. However, typical concentrations of 
simulated clusters are usually higher than $\sim3-5$ 
(Eke, Navarro, \& Frenk 1998; Brainerd, Goldberg, \& Villumsen 1998). So the 
lower mass estimates are more likely, based on theoretical expectations 
of the minimum concentrations.

There are three main sources of systematic error in the estimation
of weak lensing masses. The uncertainty about   
the actual redshifts of the background galaxies induces an uncertainty of 15-20\%
in the mass for clusters at $z\sim0.5$. In addition, the Clowe et~al.\ (2000) 
magnitude and color selection criteria for removing the red 
sequence and bright foreground galaxies did not remove the blue dwarf
cluster members. Those members dilute the shear signal by an estimated
10-20\%, increasing the mass estimate by the same amount. Also, if the dwarfs
are concentrated in the core like the giant ellipticals, this effect
lowers the measured concentration $c$ as well.
  
The best-fit weak lensing mass inside of $r_{500}$, therefore,
is $M_{500}=8.4 \times 10^{14} h^{-1} \Msun$, with a $1 \sigma$ minimum mass
of $M_{500}>3.5 \times 10^{14} h^{-1} \Msun$. The weak lensing mass increases by
$10-20\%$ if the estimated effect of dilution is taken into account. 
Even so, the weak lensing
mass is consistent with the mass derived from the velocity dispersion
and the mass derived from the X-ray data.

The third systematic in the mass conversion comes from the assumption of a
cluster shape when converting from a projected mass to the mass inside a sphere.
The weak lensing signal reports the mass in a cylinder along the line of sight,
which is converted into a mass within a spherical volume in the analysis process. 
(The masses we discuss above are masses inside a sphere.) X-ray
temperatures are affected by the mass inside a sphere. 
The full comparison of an X-ray-determined mass 
with a weak lensing mass requires the
knowledge of the cluster's intrinsic shape and mass distribution. In the
next section, we will use the SZE data to constrain the intrinsic inclination and
ellipticity of this decidedly elliptical cluster, and in the following section, we 
explore how the assumption of an ellipsoidal cluster affects our conclusions.

\section{Sunyaev-Zel'dovich Effect and Three Dimensional Shape}

We present here a joint analysis of the Sunyaev Zel'dovich Effect observations, 
the Chandra surface brightness map, and the Chandra spectral fits for 
MS0451.6-0302. 

\subsection{Sunyaev-Zel'dovich Observations}

\begin{figure}[ht]
\includegraphics[width=\textwidth]{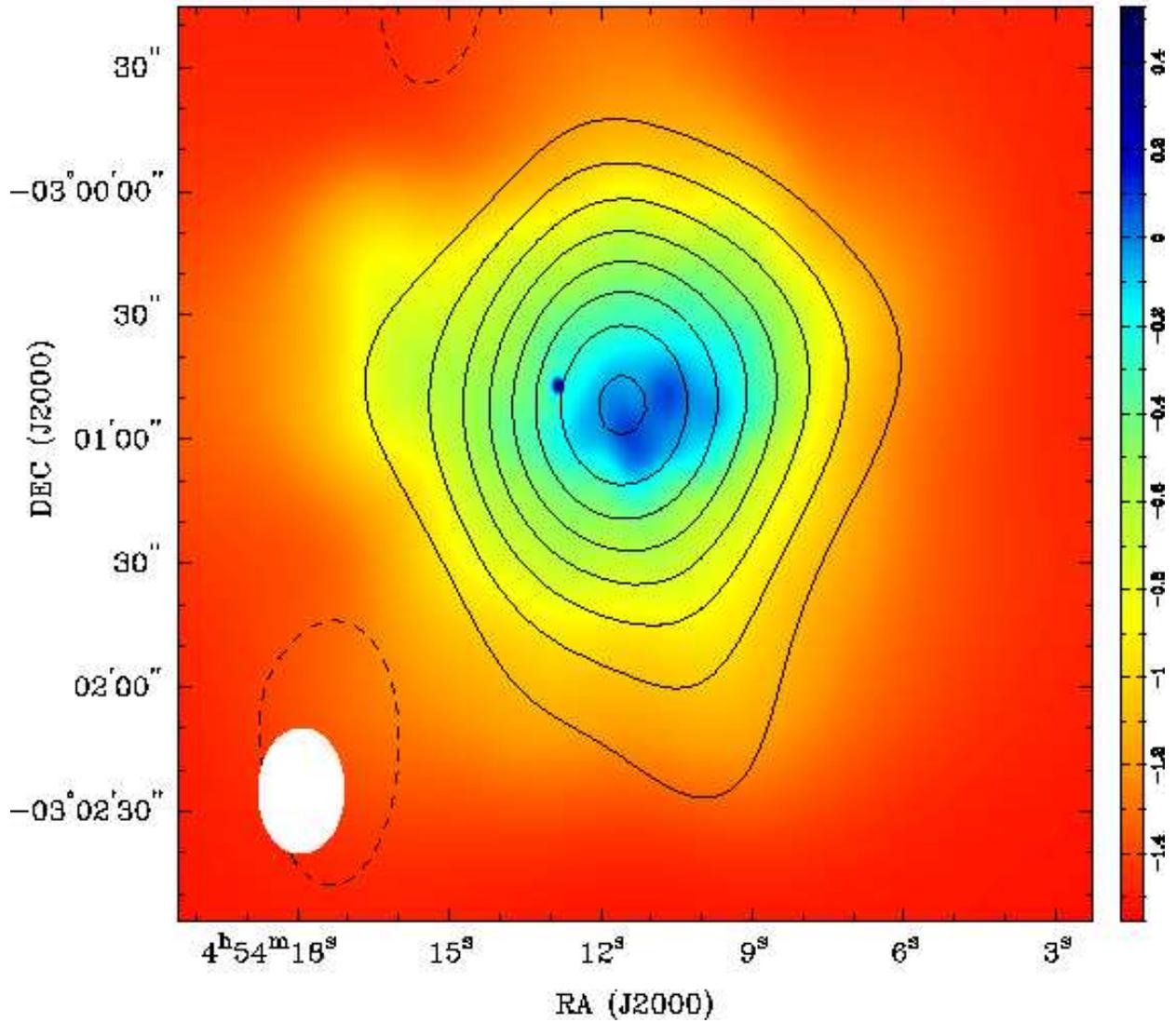}
\caption{Sunyaev-Zel'dovich Effect map, 2$\sigma$, contours for MS0451.6-0305 
superimposed on the ACIS X-ray smoothed image.
The SZE images have rms values of $\sim30~\mu$K (Reese et al. 2000). \label{figure:SZE_contours}}
\end{figure}

The SZE observations (Figure~\ref{figure:SZE_contours}) for MS0451.6-0302    
were taken at the Owens Valley Radio Observatory,
OVRO, in 1996 for a total of 30 hours using 
two 1 GHz channels centered at 28.5 GHz and 30.0 GHz as 
reported by Reese   et~al. (2000).  One point source was found in the   
cluster field and is located at 04h 54m 22s, $-03\degr \, 01\arcmin 26\arcsec$ 
(J2000).

\subsection{Joint X-ray and SZE Modeling}
The spatial parameters of the intra-cluster medium (ICM) were constrained by jointly
 fitting the Chandra X-ray spatial data, and the interferometric
 OVRO SZE data to a composite spherical $\beta-$model using the Jointfit analysis package
 developed by Reese   et~al. (2000).  The fitting algorithm simultaneously fits the source
and background X-ray models and the SZE interferometric models using a downhill
simplex method that maximizes the likelihood function 
(Cash 1979; Kendall  \&  Stuart 1979; Hughes  \&  Birkinshaw 1998). We used Jointfit
rather than SHERPA here to include the SZE data. We are encouraged that the results
from both fitting packages, using different statistical assumptions, result in
similar X-ray surface brightness parameters and uncertainties.

The parameters S$_{X0}$, $\beta$, r$_{c}$, $\Delta$T$_{0}$, and the radio point source 
fluxes (in both the 28.5GHz and 30.0 GHz channels) were allowed to vary, while the
cluster central position, point source position, and X-ray background were fixed.
 Both $\beta$ and r$_{c}$ were linked between all of the data sets,  
the SZE decrement was linked between the two SZE data sets; however, the superior
resolution of the Chandra data means that that data dominated the fits to $\beta$
and $r_c$.  Uncertainties were found by
 fixing the centroid and the point-source positions and fluxes at their best-fit 
values and then calculating the $\chi^{2}$ statistic over a large range
of S$_{X0}$, $\beta$, $r_{c}$, and $\Delta$T$_{0}$ values.   
The best-fit values and their respective uncertainties, for a 68.3$\%$ confidence 
interval (i.e. $\Delta$$\chi^{2}$=1),
 are reported in Table~\ref{tab:SZE_results}.  These values are consistent with 
 results obtained by
 Reese   et~al. (2000) using ROSAT and SZE data and with our results here using 
 Chandra data alone (Table~\ref{tab:sherpa}). 

\subsection{Electron Densities, Angular Distances, Gas Fractions, and Cluster Geometry}  

The observed  Sunyaev-Zel'dovich decrement is  $\Delta T = -1.478 \times 10^{-3}$ K. 
If the central electron density is estimated from the Sunyaev-Zel'dovich
decrement parameters (Grego et~al.\ 2001) along with only the spatial parameters from
the Chandra image (not the surface brightness normalization), 
the central electron density is
$n_e = 0.016 \pm 0.0044 h$ cm$^{-3}$, 
compared to $n_e = 0.0146 h^{1/2}$ cm$^{-3}$
from X-ray parameters alone. These two estimates differ in their dependence on $H_0$. 
Converted to $H_0=75 h_{75}$ km s$^{-1}$ Mpc$^{-1}$,
these estimates are coincident, $0.0120 \pm 0.003 h_{75}$ cm$^{-3}$ and 
$0.0126 \pm 0.0016 h_{75}^{1/2}$ cm$^{-3}$, respectively.  

Using the parameters derived from the spherical spatial SZE and X-ray and 
X-ray spectral fits, an angular diameter distance of
$D_A = 1219^{+340}_{-288}$ $^{+387}_{-387}$ Mpc 
was calculated (statistical uncertainty followed by systematic
uncertainty at 68$\%$ confidence).  The uncertainties
are reported in Table~\ref{tab:daunc}. 
The Hubble constant can be estimated from the calculated cluster distance  $D_A$ of 
MS0451.6-0321.  
Assuming an $\Omega_{M}$ = 0.3 and $\Omega_{\Lambda}$ = 0.7 cosmology, we find  
H$_{0}$ = 75$^{+23}_{-16}$ $^{+35}_{-18}$ km s$^{-1}$Mpc$^{-1}$ (statistical uncertainty 
followed by systematic 
uncertainty at 68$\%$ confidence).  

If we were to assume that the Hubble
constant derived from this single cluster is correct, $h=0.75$ and
its angular distance scale is $D_A = 1219$ Mpc. However, $H_0$ determined from
the SZE and X-ray properties of a single cluster is not particularly interesting, since
systematic uncertainties, particularly those about the cluster's geometry or
shape, are so large.
To reduce these uncertainties, it is necessary to obtain accurate spatial and spectral X-ray data 
and SZE measurements for a sample of clusters, or to precisely constrain the geometry and the
physical conditions  of an individual cluster.  Results from a sample of 18 clusters observed
with ROSAT X-ray and SZE imaging telescopes indicate
that H$_{0}$ = 60$^{+4}_{-4}$ $^{+13}_{-18}$ km s$^{-1}$ Mpc$^{-1}$ 
(Reese   et~al. 2002). 

Therefore, we take a different approach, and use the constraints on the Hubble
constant from other projects (Riess et~al.\ 1998; Freedman et~al.\ 2001) 
to determine (a) whether our assumption of
ellipsoidal gas distribution is consistent with both the X-ray and the
SZE data and (b) whether we can constrain the intrinsic axis ratio of
that distribution. 
The Hubble constant as derived from the SZE joint fit from
spherical symmetry assumptions ($H_0 \sim 75$
km s$^{-1}$ Mpc$^{-1}$) is 
somewhat higher than that obtained from 
Type Ia supernovae (Riess et~al.\ 1998), but consistent
with that of the Hubble Key Project (Freedman et~al.\ 2001).

\begin{figure}
\plotone{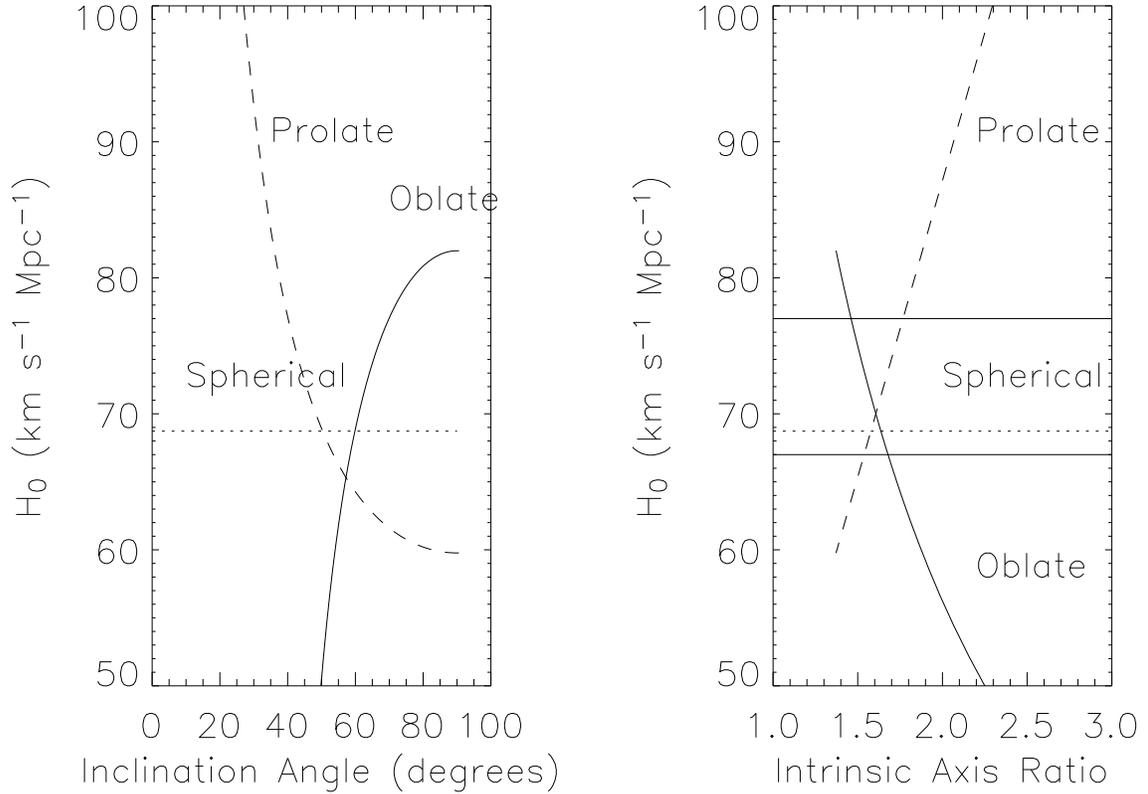}
\caption{\label{figure:geo} On both figures, we plot the inferred values of
$H_0$ for an oblate spheroid (solid line), a sphere (dotted line), and
a prolate spheroid (dashed line). On the left, the plot is a function
of inclination angle of the axis of symmetry to the line of sight. On 
the right, the plot is a function of the intrinsic ratios of the major
and minor axes. The solid horizontal lines indicate 
the 1$\sigma$ upper and lower bounds  
of $H_0 = 72 \pm 5$ km sec$^{-1}$ Mpc$^{-1}$ 
from the WMAP experiment, assuming that the universe
is flat and the distribution of fluctuations is best described by
a single power law (Spergel et al. 2003). This value is consistent with 
that found locally by the Hubble Key Project (Freedman et al. 2001)
($H_0 = 72 \pm 5 \pm 7$ km sec$^{-1}$ Mpc$^{-1}$.}
\end{figure}

We plot the inferred Hubble constant for a range of intrinsic axis ratios,
or equivalently, incidence angle $i$ in Figure~\ref{figure:geo}.
If we assume the Hubble Key Project value of 
$H_0 = 72 \pm 8$ km s$^{-1}$ Mpc$^{-1}$, the implied range of intrinsic
axis ratios for the oblate case is 1.40 -- 1.74 and for
the prolate case is 1.47 -- 1.82 (these boundaries are blurred somewhat by
the uncertainty in the observed ellipticity). 
If we assume the WMAP value (derived via an a priori assumption of a 
flat universe and a power-law fluctuation spectrum) $H_0=72 \pm 5$  
km s$^{-1}$ Mpc$^{-1}$, the implied range of intrinsic
axis ratios for the oblate case is 1.46 -- 1.67 and for
the prolate case is 1.54 -- 1.76.
With either a prolate or an
oblate geometry, there is room for significant ellipticity in the
intrinsic distribution of X-ray gas. 

However, we are not required by our data to assume extreme axis ratios.
A simple test of consistency between the X-ray data, the SZE data,
and the assumption of an ellipsoidal gas distribution show that a 
triaxial cluster model can describe the data. Thus an
extreme axis ratio, while not ruled out, 
is not necessary to explain the X-ray and SZE data.
We make this test by deriving
the core radius along the line of sight. We set 
the Hubble constant $H_0=71\pm5$ km s$^{-1}$ 
Mpc$^{-1}$ (Freedman et~al.\ 2001; Spergel et~al.\ 2003) 
and combine the central X-ray surface brightness, the
SZE temperature decrement, and the X-ray temperature values. The core 
radius along the line of sight 
derived by this procedure, assuming $\beta=0.8$ is $34\pm2\arcsec$.  
That quantity is the geometric mean  of
the best-fit spherical core radius ($30\arcsec$) and the 
best-fit elliptical core 
radius ($40\arcsec$) in the plane of the sky, 
as might be expected for a triaxial distribution of gas (i.e., 
intermediate between a prolate and an oblate distribution.) 

\begin{figure}
\plotone{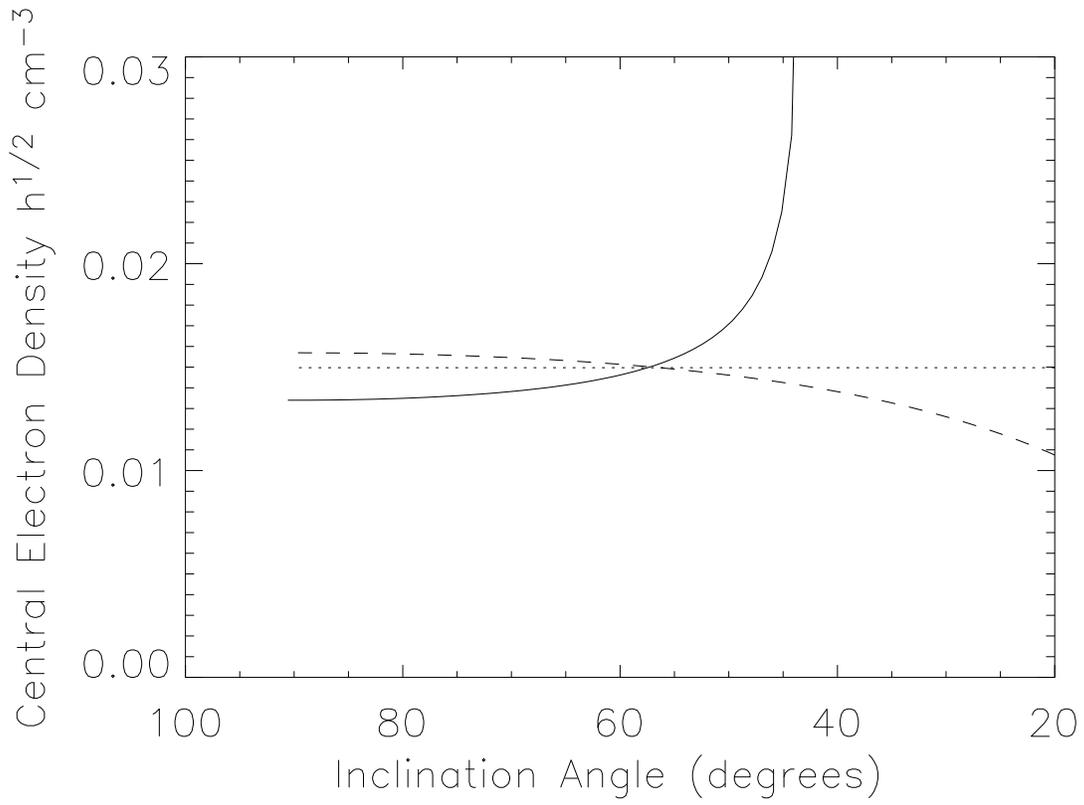}
\caption{We plot the inferred central electron density ($h^{1/2}$ cm$^{-3}$) 
and its dependence on geometry (spherical (dotted line), prolate spheroid
(dashed line), oblate 
spheroid (solid line)) and inclination angle. Note that the differences are rather small
until the inclination angle of the axis of symmetry to the line of sight gets
small ($\la 40-50\degr$. \label{figure:density_geo}) }
\end{figure}

\begin{figure}
\plotone{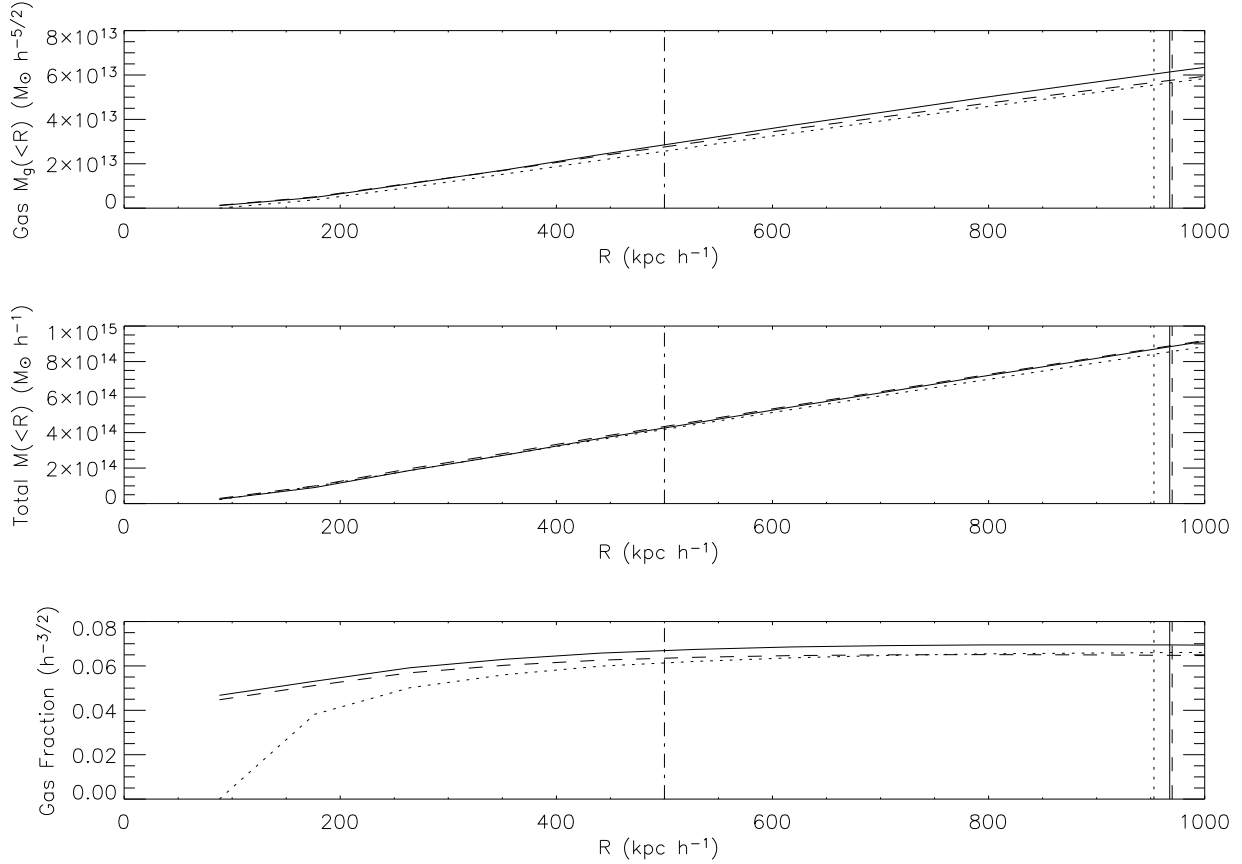}
\caption{The gas mass, total mass, and gas fraction interior to a sphere of radius
$R$ are plotted as a function of $R$ for three different shape assumptions. Beta 
parameter $\beta=0.8$ for all models, except the spherical model, which is
$\beta=0.75$. The core radius is set to the best-fit elliptical core radius 
($r_c = 180 h^{-1}$ kpc, $i=90\degr$) for the prolate (dashed line) and
oblate (solid line), and to the best-fit spherical core radius ($r_c=140 h^{-1}$ kpc)
for the spherical model (dotted line.) Note that the total mass is not very sensitive
to the assumed shape. The dot-dash vertical line indicates the outer limit
to the fit X-ray surface brightness data (at about $r_{2000}$). The vertical
lines at the far right indicate $r_{500} \sim 1 h^{-1}$ Mpc, estimated from the X-ray data,
for the various geometric assumptions. (The same line coding applies.) 
\label{figure:3mass}}
\end{figure}

The central electron densities inside $1h^{-1}$ Mpc 
as a function of assumed geometry and inclination are plotted 
in Figures~\ref{figure:density_geo}. A table of the results for $i=90\degr$ is
provided (Table~\ref{masses}).  We computed the inferred gravitational
mass distribution for the ellipsoidal geometries. The total mass inside a sphere
was not very sensitive to the assumed geometry. At $R\sim1 h^{-1}$ Mpc,
the differences between the enclosed masses were less than 1\% for models consistent
with observed cluster parameters. We plot
the gas mass, total mass, and gas fraction as a function of radius for three
geometric assumptions in Figure~\ref{figure:3mass}. On this plot, a vertical line marks
the radius ($\sim500 h^{-1}$ kpc) out to which we fit the X-ray surface brightness data.
We note that the gas fraction is relatively constant at this radius and beyond.

Recall that the ellipsoidal $\beta-$model should only be used perturbatively 
(that is, for axis ratios
not too much larger than 1.0), because it assumes that the gas is distributed on
ellipsoids which are concentric and which do not change in eccentricity or position
angle as a function of radius. For extreme axis ratios, this assumption breaks down. 
Ideally, one might want to look at cluster models generated by hydrodynamic 
numerical studies, which are beyond the scope of this paper.

\begin{figure}
\plotone{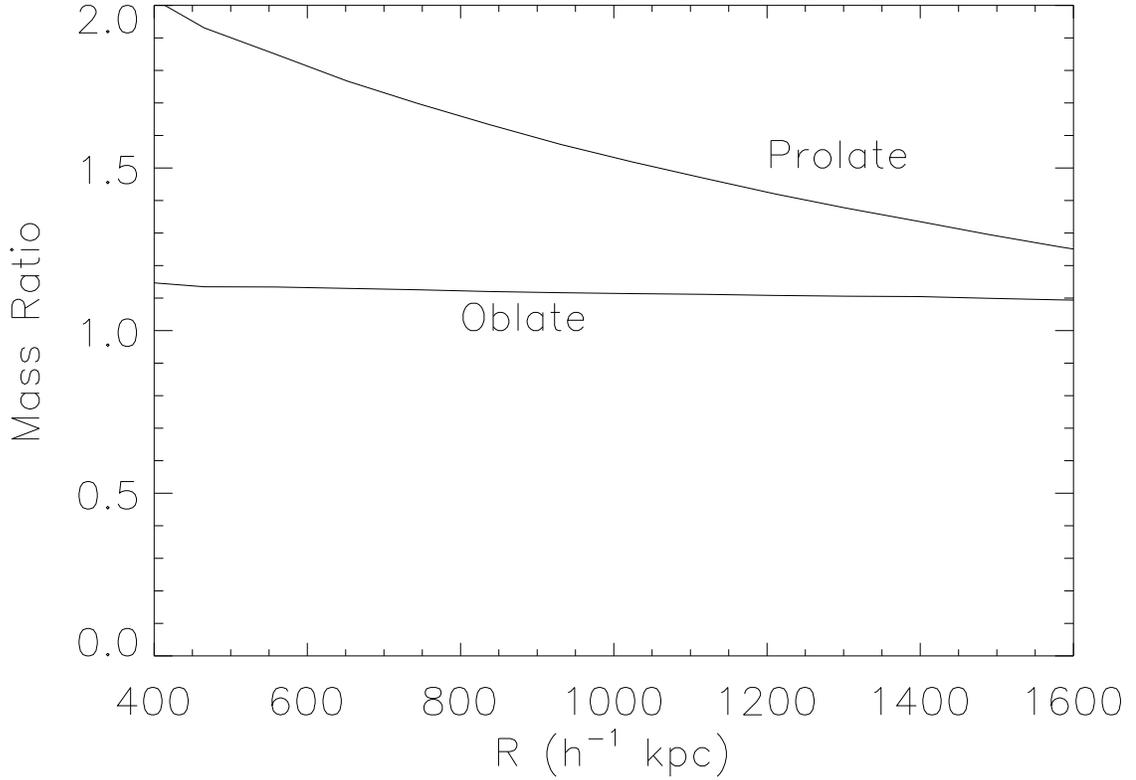}
\caption{\label{cyl_ratio} We plot as a function of radius 
the ratio of the cylindrical projected mass along a line of sight, 
truncated at the cube model borders at a distance of $\sim1.9 h^{-1}$ Mpc from the center 
of the cluster to the mass inside a sphere of the same radius. We assume an intrinsic core
radius of 190 kpc for both the prolate and oblate models, at an inclination angle of 90 
degrees. The model was computed within a volume of 
$200 \times 200 \times 200$ cells, where 1 cell is 38 kpc across, and the mass density
was projected along a cylinder perpendicular to a wall of the volume. The
correction from a spherical mass of a model to a projected mass 
depends somewhat on the assumed geometry.}
\end{figure}

We also computed the ratio of the projected gravitational mass to the
mass inside a sphere for a symmetry axis angle of 90 degrees
(Figure~\ref{cyl_ratio}). 
A range of 
ratios up to $\sim50\%$ at $1 h^{-1}$ Mpc is possible   
between the projected lensing mass and the X-ray mass between the projected
mass and the mass inside a sphere. The differences in the factor for 
correcting the projected mass to a spherical mass could be of order 40\%, 
depending on whether one assumes a prolate or an oblate model.   

\section{Discussion and Conclusions}

Using Chandra X-ray data for the cluster of galaxies MS0451.6-0321, we have
confirmed that the X-ray temperature is $(10.0-10.6)\pm^{1.6}_{1.3}$ keV (90\% confidence
range). The best-fit temperature with the current calibration still has  
an additional 0.5-1.0 keV systematic uncertainty because of the sensitivity 
to the energy bounds of the spectral
fit and the possible presence of a soft component. We also detected iron, 
at $(0.30-0.40)\pm0.14$ solar abundances. The 
unambiguous presence of iron at the level of 
present-day cluster gas metallicities confirms the lack of metallicity 
evolution in cluster gas since $z\sim 0.5-0.8$.

The cluster is decidedly elliptical in appearance. The peak and 
centroid of the X-ray surface
brightness shifts from the BCG in the soft X-ray map 
several arcseconds to the east in the hard X-ray map. We fit the surface brightness of the cluster
to spherical and elliptical $\beta-$models. The parameters of these fits were used to
derive the central electron density, the gas mass, the total mass, and the gas fraction
in the cluster as a function of radius and of assumed geometry. We explore the
the effects of the assumptions of spherically symmetric and ellipsoidal 
gas distributions.
The underlying gravitational potential was inferred from the
distribution of the hot gas assuming the gas is approximately hydrostatic
equilibrium. The 0.7--7.0 keV emission is dominated at the 95--99\% 
level by the hot component. 
The gas mass is 
somewhat sensitive to the geometry, but the total mass inside a sphere is not.
The gas fractions of $0.06-0.09 h^{-3/2}$ imply 
$\Omega_M$ of $0.3-0.2 h^{-1/2}$, if $\Omega_b = 0.019h^{-2}$. 
Using WMAP results for $\Omega_M$ and $\Omega_b$, the hot gas fraction for
this cluster (and other clusters) imply that over 2/3 of the baryons
in the universe are in between the galaxies, in an ICM or an IGM.
The Sunyaev-Zel'dovich data allows us to test the ellipsoidal assumption for
consistency, if  
$H_0$ constraints are adopted from other experiments such as the HST Key Project results,
and if the effect of gas clumpiness is minimal. We find that the data (X-ray,
SZE, and weak-lensing) are completely
consistent with a triaxial distribution of gas, intermediate between the prolate
and oblate cases. Extreme axis ratios are not necessary to explain all three
datasets. A full-fledged reconstruction of
the cluster may be possible with improved SZE, X-ray, and weak lensing data, as has been
suggested by Fox \& Pen (2002).

The overall distribution of X-ray emitting plasma in MS0451.6-0305 is ellipsoidal,  
and appears to be nearly in hydrostatic equilibrium. 
We do not see obvious signs of shocks in the hot gas. 
Shocks may have been seen in the form of linear features, or 
surface brightness enhancements, or hard (2.0--7.0 keV) features that
depart from the overall shape of the cluster. 
Elliptical surface brightness distributions themselves are consistent with  
relatively relaxed 
gravitational systems. We note our assumptions here of constant
ellipticity with radius and of concentricity  
get increasingly poor at larger radii. 
An elongated distribution of gas could result from 
filamentary, rather than spherically symmetric, infall. Infall along
filaments is predicted from numerical simulations of cluster formation.
Violent relaxation can result in a triaxial system, such as in an
elliptical galaxy. The systems may continue to interact via two-body
interaction processes, but these processes only very slowly modify the
gravitating structure of a cluster of galaxies. 

However, though the surface brightness data show an ellipsoidal distribution of gas, 
there is evidence for small departures from smoothness at the core. In particular,
multi-bandpass X-ray data as revealed by maps
of adaptively smoothed soft, mid-, and hard energy band data suggest that the 
cluster core profile may consist
of an X-ray luminous gas system surrounding the brightest cluster galaxy (BCG),
which perhaps is not yet settled into the center of the cluster.
The centroid of the hardest X-ray emission is not centered on the BCG. The residuals
from the fit of an elliptical beta-model to the surface brightness data 
reveal excesses to the south and to the east, corresponding to the
structures revealed in the color images.

We also note that the extrapolated mass at $r_{500}$ and especially at 
$r_{200}$ is sensitive to the polytropic index assumed. The ratio between
an isothermal ($\gamma=1$) mass and a $\gamma=1.2$ polytropic mass at
$r_{500} \sim 5 r_c$ (for this cluster)  was 1.8. Such a difference is of order the difference
between the theoretical mass-temperature relation (e.g.\ Evrard et~al.\ 1996)
and the observed mass-temperature relation (e.g.\ Finoguenov, Reiprich, 
\& B\"ohringer 2001; Horner, Mushotzky, \& Scharf 1999). Therefore, 
reliable temperature gradients are essential towards solving that
discrepancy.

In conclusion, we believe we have evidence that MS0451.6-0305 is not in perfect
gravitational equilibrium since there is a hint that the BCG may be just now
settling into the cluster core. However, this
interaction does not seem to be creating a violent merger shock, since the
hard X-ray image appears to be relatively symmetric, smooth, and 
single peaked, without any elongated or filametary features. The global, hot temperature
of this system, therefore, is likely to be representative of the cluster
potential.  We have, to the extent the Chandra data allows, confirmed that this
cluster is indeed as hot and massive as previous ASCA observations
(Donahue 1996; Donahue et~al.\ 1999). Additional evidence for the high mass
of this cluster is found in the velocity dispersion 
(Carlberg et~al.\ 1996), and
weak lensing data (Clowe et~al.\ 2000), which  
present mass measurements consistent with the X-ray data. 
We find some evidence for departures from equilibrium in the core of 
the cluster. These structures, resolved by Chandra but blurred by
ROSAT, mean that the surface brightness fit for this cluster 
based on ROSAT had somewhat larger, flatter core.

We find a suggestion of a soft component contributing to the emission at 
$E<0.7$ keV. This component, if real and not an artifact of 
calibration and background subtraction, is not confined to the core. 
Our data are not sufficient to distinguish a thermal from
a non-thermal spectrum for this component, as a power-law, a zero-metallicity
thermal spectrum and a 40\% solar metallicity thermal spectrum all 
adequately fit
the soft data. Deep XMM observations of
the same cluster would be useful to confirm the soft emission and to test for
the presence of 
Fe lines from $E<1$ keV thermal gas. The 2--10 keV rest luminosity of this component
is consistent with that of luminous member ellipticals or a group, similar
to what has been seen in Coma in both the ellipticals (Vikhlinin et~al.\ 2001)
and the groups and filamentary substructures (Neumann et~al.\ 2003).  This component 
does not significantly affect the
analysis of the hot phase, but it is intriguing and worth further study.

The consistency between the X-ray and Sunyaev-Zel'dovich estimates of the
central electron density and the consistency between the X-ray, weak lensing, and 
optical estimates
of the virial mass of the cluster suggest that despite the ellipsoidal
appearance of MS0451.6-0302, the global properties of this clusters are
useful for cosmological studies, independent constraints on $\Omega_M$,
and for an assessment of the hot baryon fraction in massive halos.

Most fundamentally, this Chandra study, along with a similar Chandra study for another EMSS
cluster, MS1054-0302 (Jeltema et~al.\ 2001), confirms the presence of 
high-redshift, massive clusters inside the EMSS survey volume. The 
existence of these massive clusters in a small volume is the key 
observational ingredient in studies of cluster evolution 
leading to the conclusion of a low value of
$\Omega_M$ (e.g.\ Eke et~al.\ 1998, Donahue et~al.\ 1998, 
Donahue \& Voit 1999, Henry 2000.) Such studies compare the
temperature function of clusters now with the temperature function of
clusters in the past. Since structure formation is very sensitive to 
the ambient matter density, the evolution of the mass function of clusters
is very sensitive to $\Omega_M$. Since we have confirmed that MS0451.6-0321
is indeed very massive, whether weighed using X-ray temperatures, 
optical velocities, or weak lensing signals, we have confirmed the 
high mass of this cluster, and therefore the conclusion of a 
low-density  universe. 

\acknowledgements

We acknowledge partial support from a GO grant from the Chandra X-ray Observatory
Center, GO0-1063A, and from an Hubble Space Telescope grant GO-06668.01-95. We are
grateful to G. Mark Voit for his generous review of the draft and of the
theoretical derivations and John T. Stocke for his discussion and review. 
JPH acknowledges his Chandra grant GO-1049C. 
The results in this paper are based primarily on observations made with the Chandra X-ray 
Telescope, and with data obtained from the Chandra X-ray Observatory Center, 
operated for NASA by the Smithsonian Astrophysical Observatory and 
partially based on
observations made with the NASA/ESA 
Hubble Space Telescope,  obtained from the data archive at the Space Telescope 
Science Institute. 
The HST data discussed in this paper were obtained from the 
Multimission Archive at the Space Telescope Science Institute (MAST).
The Guide Star Catalogue-II is a joint project of the Space Telescope Science Institute and the Osservatorio Astronomico di Torino. Space
Telescope Science Institute is operated by the Association of Universities for Research in Astronomy, for the National Aeronautics and Space
Administration under contract NAS5-26555. The participation of the Osservatorio Astronomico di Torino is supported by the Italian Council for
Research in Astronomy. Additional support is provided by European Southern Observatory, Space Telescope European Coordinating Facility, the
International GEMINI project and the European Space Agency Astrophysics Division.

\begin{deluxetable}{lccccc}
\centering
\tablewidth{0pt}
\tablecaption{X-Ray, Sunyaev-Zel'dovich, and Hubble Space Telescope Observations\label{tab:obs}}
\tablehead{
\colhead{Observatory} & \colhead{Instrument} & \colhead{Date Of} &
\colhead{Exposure Time} & \colhead{Frequency} & \colhead{Tracks}\\
& & Observ. & (hr) & or Wavelength  } 
\startdata
Chandra & ACIS & 2000 Oct & 11.45 & $-$ & $-$ \\
OVRO & 30 GHz SZE imager & 1996 & 30.0\enspace & $28.5,~30.0$ GHz & 8 \\
HST  &  WFPC2   & 1995 Nov & \enspace2.89 & 6895~\AA & $-$ \\
\enddata
\end{deluxetable}

\begin{deluxetable}{cccc}
\centering
\tablewidth{0pt}
\tablecaption{Best Fit Temperature, Metallicity, and $N_H$ \label{global} }
\tablehead{
\colhead{$kT$} & \colhead{$Z_\odot$}    & \colhead{$N_H$} & \colhead{$\chi^2_{red}$ (Prob)} \\
\colhead{(keV)} &\colhead{(solar)}            & $10^{22}~\rm{cm}^{-2}$
}
\startdata
\multicolumn{4}{l}{Galactic Absorption and Raymond Smith}\\ 
\tableline
$10.8^{+1.7}_{-1.4}$ & $0.34\pm 0.13$ & $0.038\pm0.018$ & 1.08 (0.17) \\
$10.1\pm 0.9$            & $0.34\pm 0.11$ & $0.05$                    & 1.07 (0.20) \\
\tableline
\noalign{\smallskip}
\multicolumn{4}{l}{Galactic Absorption and MekaL}\\ 
\tableline
$10.6^{+1.6}_{-1.3}$ & $0.40\pm 0.14$ & $0.040\pm0.017$ & 1.07 (0.20) \\
$10.0\pm 0.9$            & $0.38\pm 0.12$ & $0.05$                    & 1.07 (0.20)
\enddata
\end{deluxetable}

\begin{deluxetable}{ccccc}
\centering
\tablewidth{0pt}
\tablecaption{Energy Binning and Fit Variations \label{bin}}
\tablehead{
 Energy & kT       & $N_H$                               &  $Z_\odot$& $\chi^2_{red}$ (Prob) \\
 Range  & (keV) & $10^{22}~\rm{cm}^{-2}$ & (solar)}
\startdata
\multicolumn{5}{l}{Galactic Absorption and Raymond Smith}\\
\tableline
0.7--7.0 & $10.7\pm^{1.7}_{1.4}$ & $0.038\pm0.018$ & $0.34\pm0.13$ & 1.08 (0.17) \\
0.3--7.0 & $\phn8.5\pm0.9$ & $0.085\pm0.007$ & $0.29\pm0.09$ & 1.34 (7e-5) \\
0.5--7.0 & $\phn9.8\pm^{1.3}_{1.0}$ & $0.061\pm 0.010$ & $0.33\pm0.11$ & 1.09 (0.16) \\
0.6--7.0 & $10.2\pm^{1.4}_{1.2}$ & $0.054\pm0.014$ & $0.34\pm0.11$ & 1.08 (0.16) \\
0.7--6.0 & $11.5\pm^{2.2}_{1.7}$ & $0.034\pm0.019$ & $0.37\pm0.13$ & 1.03 (0.37) \\ 
\tableline
\noalign{\smallskip}
\multicolumn{5}{l}{Galactic Absorption and MekaL}\\
\tableline
0.7--7.0 & $10.6\pm^{1.6}_{1.3}$ & $0.040\pm0.017$ & $0.40\pm0.14$ & 1.07 (0.20) \\
0.5--7.0 & $\phn9.6\pm^{1.2}_{1.0}$  & $0.063\pm0.010$ & $0.37\pm0.12$ & 1.08 (0.16) \\
0.7--6.0 & $11.1\pm^{1.9}_{1.4}$ & $0.037\pm0.018$ & $0.41\pm0.14$ & 1.02 (0.38) \\
\enddata
\end{deluxetable}

\begin{center}
\begin{deluxetable}{lcccccc}
\rotate
\tablewidth{0pt}
\tablecolumns{6}
\tablecaption{Soft Energy Correction And Possible Soft Excess \label{tab:soft}}
\tablehead{
\colhead{Model Name} & 
\colhead{Norm\tablenotemark{a,c}} & 
\colhead{$kT$ (keV)           } & 
\colhead{$N_H (10^{22}~\rm{cm}^{-2})$ } & 
\colhead{$Z_\odot$            } & 
\colhead{$\chi^2_{red}$ (Prob)} & 
\colhead{Fit Range (keV)      } }
\startdata
1T, NH free  &$3.3 \times 10^{-3}$ & $10.2\pm^{0.9}_{1.0}$& $<0.009$ & $0.40\pm0.13$ & 1.07 (0.19) 
& 0.7--7.0\\
1T, NH free  &$3.3 \times 10^{-3}$ & $10.0\pm^{1.3}_{1.0}$& $<0.016$ & $0.39\pm0.13$ & 1.07 (0.20) 
& 0.5--7.0 \\
1T, NH fixed &$3.6 \times 10^{-3}$ & $ 7.9\pm^{0.7}_{0.6}$& 0.05     & $0.33\pm0.09$ & 1.19 (0.02) 
& 0.7--7.0 \\
1T, NH fixed &$3.7 \times 10^{-3}$ & $ 7.4\pm^{0.5}_{0.5}$& 0.05     & $0.31\pm0.08$ & 1.25 $(3.7\times10^{-3})$ & 0.5--7.0 \\
2T, NH free  &$3.2 \times 10^{-3}$ & $10.6 (>8.0)$        & $<0.022$ & $0.42\pm^{0.17}_{0.14}$ & 1.08 (0.184) & 0.5--7.0 \\
&$2.4 \times 10^{-4}$ & $2.7$ \tablenotemark{b} &        &                         &   & \\
2T, NH free  &$3.1 \times 10^{-3}$ & $10.9 (>9.5)$        & $0.04\pm^{0.13}_{0.04}$ & $0.42 (>0.28)$ & 1.08 (0.19) & 0.5--7.0 \\
$+$ metals free     &$1.8 \times 10^{-3}$  & $ 0.7$ \tablenotemark{b} & &        $0.0$ \tablenotemark{b} &   &  \\
2T, NH free   &$3.3 \times 10^{-3}$ & $10.2\pm^{1.5}_{1.2}$& $<0.04$  & $0.40\pm0.14$ & 1.08 (0.18) 
& 0.7--7.0\\
$+$ metals equal    &$8.3 \times 10^{-5}$ & $0.7$ \tablenotemark{b}&          &               &              \\
2T, NH fixed  &$3.3 \times 10^{-3}$ & $10.1\pm^{1.6}_{1.1}$& 0.05     & $0.40\pm0.14$ & 1.08 (0.15) 
& 0.7--7.0 \\
$+$ metals equal    &$5.9 \times 10^{-4}$  & $0.7$ \tablenotemark{b} &        &               &             \\
1T+PL, NH fixed &$2.9 \times 10^{-3}$ & $11.4\pm^{4.1}_{1.8}$ & 0.05 & $0.50\pm^{0.54}_{0.17}$ & 1.07 (0.186) & 0.5--7.0 \\
		      &$1.1 \times 10^{-4}$  & $2.6$ \tablenotemark{c} &      &    &   & \\
1T+PL, NH free &$2.8 \times 10^{-3}$ & $11.1\pm^{1.5}_{2.5}$ & $<0.06$ & $0.51\pm^{0.87}_{0.23}$ & 1.08 (0.183) & 0.5--7.0 \\
                      &$9.6 \times 10^{-5}$  & $2.1$ \tablenotemark{c} &      &    &  & \\
\enddata
\tablenotetext{a}{Normalization is given in units of the X-ray emissivity integral:
$10^{-14} / [4 \pi (D_A (1+z))^2] \int n_e n_H dV$ where electron density ($n_e$) and hydrogen 
density ($n_H$) are in units cm$^{-3}$, and the angular distance $D_A$ and $dV$ are in units of
cm and cm$^3$, respectively.}
\tablenotetext{b}{The normalization, temperature, and (for some models) the metallicity of the soft component 
are the best fits; but they are very poorly constrained by the data and  
are sensitive to the bandpass of the fit. }
\tablenotetext{c}{Power-law (PL) normalization and photon-index are  best fit
quantities, but they 
are not well constrained by the data. The power law normalization is quoted 
in units of photons keV$^{-1}$ cm$^{-2}$ s$^{-1}$ at 1 keV.}
\end{deluxetable}
\end{center}

\begin{deluxetable}{lllcl}
\centering
\tablewidth{0pt}
\tablecaption{Comparison of Spectral Results for Inner and Outer Regions\label{inner_outer}}
\tablehead{
\colhead{Aperture} & \colhead{Model} & \colhead{kT} & \colhead{$Z_\odot$} &\colhead{$\chi^2_{red}$ (Prob)}
} 
\startdata
Inner	& MekaL & 10.3 (8.8-12.5) & 0.39 (0.22-0.57) &  1.3 (2.1e-3)  \\
Inner   & Raymond & 10.7 (9.0-13.2) & 0.35 (0.19-0.51) & 1.3 (1.9e-3) \\
Outer   & MekaL & \phn9.8 (8.1-12.3) & 0.26 (0.09-0.44) & 1.1 (0.14) \\
Outer   & Raymond & \phn9.9 (8.2-12.4) & 0.23 (0.08-0.40) & 1.1 (0.14) \\
BCG & MekaL & 10.7 (8.0-16.5) & 0.39 (0.07-0.72) & 1.3 (3.3e-2) \\
BCG & Raymond & 11.2 (8.2-17.5) & 0.33 (0.02-0.64) & 1.3 (3.3e-2) \\
\enddata
\end{deluxetable}

\begin{deluxetable}{lcc}
\centering
\tablewidth{0pt}
\tablecaption{SHERPA 2-D Fit Results for MS0451\label{tab:sherpa}}
\setlength{\tabcolsep}{6pt}
\tablehead{
\colhead{Parameter} & \colhead{Derived Value\tablenotemark{a}} & \colhead{Units}}
\startdata
\multicolumn{3}{c}{Spherical Beta Model, Exposure Correction }\\
\tableline
\noalign{\smallskip}
$\beta_{sph}$   & $0.70 \pm0.07$ &         \\
$r_{c,sph}$     & $30.7 \pm 3.5$\phn & arcsec \\
$S_{X0,sph}$	& $(5.58 \pm 0.23) \times 10^{-4} $ & photons~s$^{-1}$~cm$^{-2}$~arcmin$^{-2}$ \\
Bkgd (spherical) & $(4.2 \pm^{2.8}_{11.2}) \times 10^{-6}$ &  photons~s$^{-1}$~cm$^{-2}$~bin$^{-1}$ \\ 
\cutinhead{Spherical Beta Model, No Exposure Correction }
$\beta_{sph}$ (no exp) & $0.67 \pm 0.07$ &        \\ 
$r_{c,sph}$ (no exp) & $28.3 \pm 3.3$\phn & arcsec \\
$S_{X0,sph}$    & $ 0.336 \pm 0.016$			  & counts~arcmin$^{-2}$~s$^{-1}$ \\
Bkgd (spher,no exp) & $( 2.0 \pm^{1.8}_{5.7} ) \times 10^{-3}$ & counts~arcmin$^{-2}$~s$^{-1}$          \\ 
\cutinhead{Ellipsoidal Beta Model, Exposure Correction}
$\beta_{ell}$        & $ 0.79 \pm 0.08$ &                            \\
$r_{c,ell}$     & $ 40.2 \pm 4.1$\phn &  arcsec                     \\
$\epsilon_{SHERPA}$ & $0.271 \pm 0.016$ & \\
Observed axis ratio ($e_{HB}$) & $1.37 \pm 0.03 $ & \\
Position Angle (E from N) & $105 \pm 2$\phn\phn & degrees \\
$S_{X0,ell}$    & $ (5.48 \pm 0.23) \times 10^{-4}$ & photons~s$^{-1}$~cm$^{-2}$~arcmin$^{-2}$ \\
Bkgd (ell)       & $( 7.3 \pm 2.3) \times 10^{-6}$ &  photons~s$^{-1}$~cm$^{-2}$~arcmin$^{-2}$ \\
\cutinhead{Ellipsoidal Beta Model, No Exposure Correction }
$\beta_{ell}$ (no exp) & $ 0.75 \pm 0.07 $ &                            \\
$r_{c,ell}$     & $ 37.8 \pm ^{4.1}_{3.8}$\phn &  arcsec                     \\
$\epsilon_{SHERPA}$ & $0.276 \pm 0.016$ & \\
Observed axis ratio ($e_{HB}$) & $1.38 \pm 0.03 $ & \\
Position Angle (E from N) & $100 \pm 2$\phn\phn & degrees \\
$S_{X0,ell}$    & $ 0.332 \pm 0.016$ & counts~s$^{-1}$~arcmin$^{-2}$ \\
Bkgd (ell, no exp) & $(2.0 \pm^{1.8}_{5.7}) \times 10^{-3}$ & counts~arcmin$^{-2}$~s$^{-1}$          \\ 
\enddata
\tablenotetext{a}{90\% projected uncertainties are quoted for all estimated parameters.}
\end{deluxetable}

\begin{deluxetable}{cc}
\centering
\tablewidth{0pt}
\tablecaption{Cluster X-ray Flux and Luminosity \label{tab:lum}}
\tablehead{\colhead{Bandpass (keV)}     & \colhead{Flux ($\flux$)}} 
\startdata
0.5--2.0             & $9.24 \times 10^{-13}$  \\
0.7--7.0             & $2.34 \times 10^{-12}$  \\ 
\noalign{\smallskip}
\tableline
\noalign{\smallskip}
&Luminosity ($\lum h^{-2}$) \\  
\noalign{\smallskip}
\tableline
\noalign{\smallskip}
\phn2--10    & $1.05 \times 10^{45}$ \\
0.1--2.4 & $4.60 \times 10^{44}$  \\
Bolometric & $2.01 \times 10^{45}$ \\
\enddata
\end{deluxetable}

\begin{deluxetable}{lcc}
\centering
\tablewidth{0pt}
\tablecaption{Derived Quantities from X-ray Measurements\tablenotemark{a} \label{xray_masses}}
\tablehead{\colhead{Quantity} & \colhead{Estimated Value\tablenotemark{b}} & \colhead{Units}}
\startdata
Central Density $n_0$ &  $0.0146\pm0.002$ & $h^{1/2}$ cm$^{-3}$ \\
$M_{gas}$ & \phn\phn(5.6--$6.2)\pm 3.3 \times 10^{13}$ & $h^{-5/2}~\Msun$ \\
$M_{tot}$ & \phn\phn(8.6--$8.9)\pm 1.2 \times 10^{14}$ & $h^{-1}~\Msun$ \\
$f_{gas}$ & (0.065--$0.069) \pm 0.01$\phn\phn\phn\phn\phn\phn\phn  & $h^{-3/2}$ \\
$\Omega_M h^{1/2}$ & $\leq 0.29\pm0.05$\phn 
\enddata
\tablenotetext{a}{Gas mass, total mass, and gas fraction are reported here inside
$r_{500} \sim 0.97 h^{-1}$ Mpc. See Table~\ref{masses} for masses inside a metric
radius of $1 h^{-1}$ Mpc.}
\tablenotetext{b}{Systematic and statistical errors are combined in quadrature,
as in Patel et al. (2000). }
\end{deluxetable}

\begin{center}
\begin{deluxetable}{lcc}
\tablecaption{Maximum-Likelihood Jointfit Results for MS0451\label{tab:SZE_results}}
\tablewidth{0pt}
\setlength{\tabcolsep}{6pt}
\tablehead{
\colhead{Parameter} & \colhead{Derived Value} & \colhead{Units}
 }
\startdata
$\beta$        & $ 0.780 \pm ^{0.028}_{0.025}    $ &                            \\
$r_{c}$     & $ 33.95 \pm ^{1.7}_{1.6}$\phn &  arcsec                     \\
$S_{X0}$       & $ 0.346 \pm ^{0.01}_{0.01}  $ &  counts~arcmin$^{-2}$~s$^{-1}$          \\
Bkgd         &  $4.57 \times 10^{-3} \pm ^{0.0002}_{0.0004}$\phn\phn\phn\phn &  counts~arcmin$^{-2}$~s$^{-1}$ \\

$\Delta T_{0}$ & $ -1478 \pm ^{118}_{102}\phn     $ &  $\mu$~K                            \\

Point~Source~Flux$@28.5~GHz$        & \phn$0.53 \pm ^{0.1}_{0.1} $ &  mJy                    \\
Point~Source~Flux$@30.0~GHz$        & \phn$0.38 \pm ^{0.1}_{0.1} $ &  mJy                    \\
\enddata
\end{deluxetable}
\end{center}

\begin{deluxetable}{lc}
\centering
\tablecaption{Uncertainty in Cluster Distance\label{tab:daunc}}
\tablewidth{0pt}
\tablehead{
\colhead{Source of Error} & \colhead{$\delta$ $D_A$ (Mpc)}} 
\startdata
\underline {Statistical Uncertainty}   &                        \\        
                                       &                        \\
X-ray/SZE Spatial Model\tablenotemark{a} & $+295,-247$          \\ 
X-ray Spectral Model\tablenotemark{b}  & $+168,-148$            \\
30 GHz Point Source Flux\tablenotemark{f} & $+7,-7$                \\ 
& \underline {$~~~~~~~~~~~~~~~~~$}                              \\
Combined Statistical Uncertainty       & $+340,-288$            \\ 
\tableline
                                       &                        \\
\underline {Systematic Uncertainty}    &                        \\
                                       &                        \\
Absolute X-ray Flux Calibration\tablenotemark{d} & \phn$\pm61$                \\ 
Absolute SZE Flux Calibration\tablenotemark{f} &  \phn$\pm98$                \\
Undetected 28.5 GHz Point Sources\tablenotemark{e} & $\pm195$               \\
Peculiar Velocity\tablenotemark{c}     & \phn$\pm98$                \\ 
Asphericity\tablenotemark{e}           & $\pm171$               \\ 
Clumping \& Thermal Structure\tablenotemark{e} & $\pm244$               \\ 
 & \underline {$~~~~~~~~~~~~~~~~~$}       			\\
Combined Systematic Uncertainty        & $+387,-387$            \\ 
\enddata
\tablenotetext{a}{Statistical error ($68\%$ confidence interval) due to variations in the spatial model over allowed values of $\theta_c$, $\beta$,$S_{x0}$, and $\Delta T_0$.}
\tablenotetext{b}{Statistical uncertainty in modeling x-ray spectral data (68\% confidence interval).}
\tablenotetext{c}{Assuming a line of sight peculiar velocity of $v_r = \pm300$ km s$^{-1}$ 
(Reese   et~al. 2002).}
\tablenotetext{d}{Elsner   et~al. (2000); Schwartz   et~al. (2000)}
\tablenotetext{e}{Reese   et~al. (2000)}
\tablenotetext{f}{Reese   et~al. (2002)}
\end{deluxetable}

\begin{deluxetable}{lccc}
\centering
\tablewidth{0pt}
\tablecaption{Dependence of Derived Quantities from X-ray Measurements on Geometry \label{masses}}
\tablehead{\colhead{Quantity} & \colhead{Assumption} & \colhead{Estimated Value} & \colhead{Units}}
\startdata
Central Density $n_0$ & Spherical & $0.0146$ & $h^{1/2}$ cm$^{-3}$ \\
Central Density $n_0$ & Prolate, $i=90\degr$ & $0.0159$& $h^{1/2}$ cm$^{-3}$ \\
Central Density $n_0$ & Oblate, $i=90\degr$  & $0.0135$& $h^{1/2}$ cm$^{-3}$ \\
$M_{gas}$ ($R< 1h^{-1}$ Mpc) & Spherical & $5.8 \times 10^{13}$ & $h^{-5/2}~\Msun$ \\
$M_{gas}$ ($R< 1h^{-1}$ Mpc) & Prolate ($i=90\degr$)   & $5.9  \times 10^{13}$ & $h^{-5/2}~\Msun$ \\
$M_{gas}$ ($R< 1h^{-1}$ Mpc) & Oblate ($i=90\degr$)    & $6.3  \times 10^{13}$ & $h^{-5/2}~\Msun$ \\ 
$M_{tot}$ ($R< 1h^{-1}$ Mpc) & Spherical & $8.8 \times 10^{14}$ & $h^{-1}~\Msun$ \\
$M_{tot}$ ($R< 1h^{-1}$ Mpc) & Prolate ($i=90\degr$) & $9.2 \times 10^{14}$ & $h^{-1}~\Msun$ \\
$M_{tot}$ ($R< 1h^{-1}$ Mpc) & Oblate ($i=90\degr$)  & $9.1 \times 10^{14}$ & $h^{-1}~\Msun$ \\
\enddata
\end{deluxetable}

\clearpage

\end{document}